\documentclass[aps,prl,reprint,showpacs,superscriptaddress,longbibliography]{revtex4-1}
\usepackage{mathtools,amssymb,graphicx,units}
\usepackage[plainpages=false,pdfpagelabels,colorlinks=true,linkcolor=red,urlcolor=blue,citecolor=blue,pdftitle={Title},pdfauthor={},pdfdisplaydoctitle=true,pdfduplex=DuplexFlipLongEdge]{hyperref}
\usepackage{verbatim}
\usepackage{subfigure}
\usepackage[english]{babel}
\usepackage{color}
\usepackage[section]{placeins}
\usepackage{hyperref}
\usepackage[normalem]{ulem}
\usepackage{bm}
\makeatletter

\newcommand{\Rmnum}[1]{\expandafter\@slowromancap\romannumeral #1@}
\makeatother
\hypersetup{
colorlinks = true,
linkcolor = [rgb]{0.70,0.13,0.13},
citecolor = [rgb]{0.13,0.55,0.13},
urlcolor  = [rgb]{0.25, 0.41, 0.88}}
\newcommand{\ket}[1]{|#1\rangle}

\begin{document}

\title{Exploring unconventional quantum criticality in the $p$-wave-paired  Aubry-Andr\'{e}-Harper model}
\author{Ting Lv}
\affiliation{College of Physics, Nanjing University of Aeronautics and Astronautics, Nanjing, 211106, China}
\affiliation{Key Laboratory of Aerospace Information Materials and Physics (NUAA), MIIT, Nanjing 211106, China}

\author{Yu-Bin Liu}
\affiliation{College of Physics, Nanjing University of Aeronautics and Astronautics, Nanjing, 211106, China}
\affiliation{Key Laboratory of Aerospace Information Materials and Physics (NUAA), MIIT, Nanjing 211106, China}

\author{Tian-Cheng Yi}
\affiliation{Beijing Computational Science Research Center, Beijing 100193, China}

\author{Liangsheng Li}
 \affiliation{Science and Technology on Electromagnetic Scattering Laboratory, Beijing 100854, China}

 \author{Maoxin Liu}
\affiliation{School of science, Beijing University of Posts and Telecommunications, Beijing 100876, China}

\author{Wen-Long You} \email{wlyou@nuaa.edu.cn}
\affiliation{College of Physics, Nanjing University of Aeronautics and Astronautics, Nanjing, 211106, China}
\affiliation{Key Laboratory of Aerospace Information Materials and Physics (NUAA), MIIT, Nanjing 211106, China}

\begin{abstract}
We have investigated scaling properties near the quantum critical point between the extended phase and the critical phase in the Aubry-Andr\'{e}-Harper model with $p$-wave pairing, which have
rarely been exploited as most investigations focus on the localization transition from the critical phase to the localized phase.  We find that the spectrum averaged entanglement entropy and the generalized fidelity susceptibility act as eminent universal order parameters of the corresponding critical point without gap closing. We introduce a Widom scaling ansatz for these criticality probes to develop a unified theory of critical exponents and scaling functions.
We thus extract the correlation-length critical exponent $\nu$ and the dynamical exponent $z$ through the finite-size scaling given the system sizes increase in the Fibonacci sequence. The retrieved values of 
$\nu \simeq 1.000$ and 
$z \simeq 3.610$ indicate that the transition from the extended phase to the critical phase belongs to a different universality class from the localization transition. Our approach sets the stage for exploring the unconventional quantum criticality and the associated universal information of quasiperiodic systems in state-of-the-art quantum simulation experiments.
\end{abstract}
\date{\today}
\maketitle

\section{Introduction}
\label{Sec1:Introduction}
The concept of disorder induced localizations was originally addressed in the seminal work of Anderson~\cite{Anderson1958Absence} and still fueled a variety of new ideas and research
directions. Notably, engineering random disorder remains an experimental challenge, and the theoretical study of randomly disordered system suffers from the necessity of disorder averaging. It is largely recognized that there is another class of systems in which the distribution of the disordered potential is not random, but showcases some quasiperiodic structures.
Quasiperiodic disorder provides a midway between randomly disordered and clean systems for exploring novel phases of matter.
The physics of quasiperiodic systems is known to show unconventional phenomena including mobility edges~\cite{Luschen2018Single,Yao2019Critical,Wang2020One,Biddle2011Localization}, fractal bands~\cite{Liu2021Multichannel,Tanese2014Fractal},
many-body localization~\cite{Nicolas2019Many-body,Michal2014Delocalization,Mastropietro2015Localization,Znidaric2018Interaction}, topological features~\cite{Kitagawa2010Topological,Thakurathi2013Floquet,Tang2022Topological},
and exotic forms of phases~\cite{Yao2020Lieb,Bohlein2012Experimental} as well as currently intensively investigated non-Hermiticities~\cite{Weidemann2022TopologicalTP}.
Remarkably, quasiperiodic systems have been observed in a number of systems, such as photonic crystals~\cite{Negro2003Light,Lahini2009Observation}, cavity polaritons~\cite{Degottardi2013Majorana}, cold atoms in bichromatic laser potentials~\cite{Modugno2010Anderson,Roati2008Anderson,Schreiber2015Observation,Bordia2017Periodically}, twisted bilayer graphene~\cite{deDios2020Magnetoconductivity}, and optical waveguides~\cite{Pertsch2004Nonlinearity}.
The continuing development of
experimental techniques
has led to a deeper understanding of quasiperiodic criticality~\cite{Agrawal2020Universality,Agrawal2020Quantum,Chandran2017Localization,Crowley2018Quasiperiodic}
and associated universal information~\cite{Khemani2017Two,Szabo2018Non,Hida2004New,Goblot2020Emergence}.

Among a diversity of quasiperiodic models, the Aubry-Andr\'{e}-Harper (AAH) model~\cite{Harper1955General,Kohmoto1983Localization,Hiramoto1989Scaling} and its generalizations are nowadays drawing increasing attention. The standard AAH model can be formally derived from a tight-binding
square-lattice Hamiltonian in the presence of a magnetic
field yielding the famous Hofstadter butterfly to a one-dimensional (1D) chain when the hopping amplitudes are the same~\cite{Hofstadter1976Energy}.
A dual transformation between the coordinate and momentum spaces will yield the same Hamiltonian with hopping and potential amplitudes interchanged. As a consequence of the inherent self-duality,
all the eigenstates of the system undergo 
Anderson localization transition from an extended phase (EP) to a localized phase (LP) at a critical value of the incommensurate potential strength~\cite{Wei2019Fidelity,Sinha2019Kibble}, as was
first shown by Aubry and Andr\'{e}~\cite{Aubry1980Analyticity}.
By contrast, all single-particle states localize for arbitrarily weak disorder
in 1D systems with uncorrelated disorder~\cite{Anderson1958Absence}.
During the past few years, a growing effort has been
made to explore the effect of the self-duality breaking interactions to the AAH Hamiltonian. A peculiar direction is the inclusion of the $p$-wave superconducting pairing~\cite{Zeng2016Generalized,Cai2013Topological,Wang2016Spectral}.
The reentrant localization transition can be established by analyzing inverse participation ratios (IPRs) of eigenspectra, the bandwidth distribution and the level spacing distribution.
The $p$-wave pairing term breaks the self-dual symmetry and splits the single transition into a three-phase spectral diagram.
A critical phase (CP) crops up between the extended and localized phases~\cite{wang2016Many-body,Wang2016Phase,Tong2021Dynamics}.
The emerging CP exhibits various interesting features, such as power-law localization~\cite{Satija1993Symmetry}, the critical spectral statistics~\cite{Geisel1991New,Machida1986Quantum,Bertrand2016Anomalous}, multifractal behavior of wave functions~\cite{Mirlin2006Exact,Dubertrand2014Two}.
In particular, the intermediate CP proliferates in the quantum magnetism described by the anisotropic XY chain
under an irrationally modulated
transverse field, which is equivalent to the AAH model with $p$-wave pairing through the celebrated Jordan-Wigner transformation~\cite{Katsura1962Statistical,Satija1989Localization,Doria1988Quasiperiodicity,Satija1994Spectral}.
The CP can be possibly detected by measuring the mean square displacement of the wave packet after a fixed time of expansion 
in the real space or 
the momentum distributions~\cite{Wang2020Realization}.

While most investigations focus on the phase transition between the CP and the LP, little is known about critical properties of the transition from the EP
to the CP. The critical-localized phase transition is deemed as the reminiscence of Anderson localization transition in the absence of $p$-wave pairing, which
coincides a second-order transition from a topological superconducting phase to a topologically trivial localized phase~\cite{Cai2013Topological}.
The quench dynamics from the LP to the CP through the Kibble-Zurek mechanism and the gap scaling unveil the correlation-length exponent $\nu \simeq 1.000$ and the dynamical critical exponent $z \simeq 1.373$~\cite{Tong2021Dynamics}, which are consistent with numerical results of the generalized fidelity susceptibility (GFS) $\nu \simeq 1.000$, $z \simeq 1.388$~\cite{Lv2022Quantum}.
However, the gap scaling fails owing to the apparent nonclosure of the spectral gap across the extended-critical transition with periodic boundary conditions, which is beyond the regime of thermodynamic phase transition.
Therefore, there is a clear need to identify observables for the characterization of the EP-CP transition that can be measured in state-of-the-art quantum simulators and are easy to compute numerically as well.

Motivated by these open questions, in this work, we bring in new tools to understand the
nature of extended-critical transitions. The rest of the paper is organized as follows. We briefly introduce the AAH Hamiltonian with $p$-wave pairing in Sec.~\ref{sec:model}. Section~\ref{sec:vNE} introduces the spectrum averaged von Neumann entropy. We then provide evidence that such multipartite entanglement measure is well suited for
characterizing the extended-critical transition and the fractal dimension of the critical phase.
Section~\ref{sec:GFS} is devoted to the finite-size scaling (FSS) of GFSs,
and extracting the critical exponents that cannot be obtained from conventional scaling analysis.
In Sec.~\ref{sec:SUMMARY}, we give a brief summary.

\section{MODEL HAMILTONIAN}\label{sec:model}
We consider the Hamiltonian for the AAH model with $p$-wave pairing in a quasi-periodically modulated potential,
\begin{eqnarray}
H\!=\!\sum_{j=1}^N\!(-Jc_{j}^{\dag}c_{j+1}+\Delta c_{j}c_{j+1}+{\rm H.c.})\!+\!
\sum_{j=1}^N V_{j}(c_{j}^{\dag}c_{j}-\frac{1}{2}),~~~~
\label{Ham}
\end{eqnarray}
where $c_{j}^{\dag}$ $(c_{j})$ is the fermionic creation (annihilation) operator at site $j$ among total $N$ lattice sites, $J$ denotes the nearest-neighbor hopping amplitude, $\Delta$ characterizes the strength of $p$-wave superconducting pairing, and H.c. stands for the Hermitian conjugate. In certain context, the superconducting order parameter may appear in the mean-field approximation of interacting AAH model~\cite{Zhang2022Enhanced}.
Without losing generality, we assume $J=1$ as an energy unit throughout the paper.
The external potential in Eq. (\ref{Ham}) is quasiperiodic, i.e., $V_{j}=V\cos(2\pi\alpha j+\phi)$, where $\phi \in [0,2\pi)$ is a random phase and $V$ is the amplitude of the potential
with an irrational wave number $\alpha$.
A commonplace choice for $\alpha$ is the inverse golden ratio $\alpha=(\sqrt{5}-1)/2$. The boundary condition is imposed as $c_{N+1}$=$\sigma$$c_{1}$, where $\sigma$= $1$, $-1$, and $0$ corresponding to periodic, antiperiodic, and open boundary conditions, respectively.
The Hamiltonian describes the Kitaev $p$-wave
superconducting model for $\alpha= 0$, while the model reduces to the celebrated Aubry-Andr\'{e} model when $\Delta=0$,
which undergoes a localization-delocalization transition at $V=2J$~\cite{Harper1955General,Kohmoto1983Localization,Hiramoto1989Scaling}.
Once the $p$-wave pairing is turned on, the symmetry breaking from U(1) down to ${\mathbb Z}_2$ leads to the emergence of the CP sandwiched between these two phases~\cite{Wang2016Phase,liu2017Phase,Tong2021Dynamics}.
Upon increasing the Aubry-Andr\'{e} potential
strength $V$, the system undergoes continuous transitions from the CP to the EP for $V_{c1}=2|\Delta-J|$ and
from the CP to the LP for $V_{c2}=2|\Delta+J|$~\cite{Wang2016Spectral,Wang2016Phase}.
Analogous to the case of $\Delta=0$, all the eigenstates of the Hamiltonian in Eq.(\ref{Ham}) with a generic $\Delta$ undergo two transitions simultaneously. For $\Delta=\pm 1$, the model will be equivalent to quasiperiodic Ising model~\cite{Fisher1995Critical,Chandran2017Localization},
in which the phase transition occurs only between critical and localized phases~\cite{Satija1993Symmetry,Satija1994Spectral}.
The FSS of generalized participation indicates that the correlation-length exponent $\nu = 1$ across both transitions for all irrational $\alpha$~\cite{Wang2016Spectral},
consistent with the Harris criterion~\cite{Harris1974Effect}, which imposes that
$\nu < 2$ for phase transitions in the presence of incommensurate modulation.
In the numerical treatment on finite lattices, it is convenient to replace the inverse golden ratio with a rational approximant being a ratio of two Fibonacci numbers, $\alpha=F_{\ell-1}/ F_{\ell}$, and thus the diagonal Aubry-Andr\'{e} potential has periodicity of $F_{\ell}$ sites in order to allow for the use of periodic boundary conditions, where $F_{\ell}$ is the ${\ell}$-th Fibonacci number. As noted before~\cite{Ino2006Critical,Gong2008Fidelity,Cookmeyer2020Critical},
the sequence of denominators $F_{\ell}$ breaks into three subsequences.
As we will show, each subsequence is characterized with a separate scaling function.
The lattice system $N$ can be chosen as $\zeta$ supercells, where $N = \zeta F_{\ell}$. For simplicity, we consider $\zeta=1$ and two-subsequence odd sizes in the following, i.e., $F_{3\ell+1}= 21, 55, 233, 987, \ldots$, and $F_{3\ell+2}= 89, 377, 1597, 6765,\ldots$.
In this respect,
the Hamiltonian (\ref{Ham}) can be diagonalized through a canonical Bogoliubov-de Gennes (BdG) transformation by introducing quasiparticle operators $\eta_{k}$ and $\eta_{k}^{\dag}$, which is a linear combination of an electron and hole:
\begin{eqnarray}
\eta_{k}&=&
\sum_{j=1}^{N}(u_{k,j}c_{j}+v_{k,j} c_{j}^{\dag}),\quad
\eta_{k}^{\dag}=
\sum_{j=1}^{N}(u_{k,j}^*c_{j}^{\dag}+v_{k,j}^* c_{j}), \quad \quad
\end{eqnarray}
where $u_{k,j}$ and $v_{k,j}$ denote electron and hole amplitudes  of Bogoliubov quasiparticle
at site $j$ for $k$-th eigenstate. For a given normalized wave
$({\sum_{j}\vert u_{k,j}\vert^{2}+\vert v_{k,j}\vert^{2}=1})$, $\eta_{k}$ and $\eta_{k}^{\dag}$ satisfy the anticommutation relation:
\begin{eqnarray}
\{\eta_{k},\eta_{k'}^{\dag}\}&=&
\delta_{kk'},\quad
\{\eta_{k},\eta_{k}\}=
\{\eta_{k}^{\dag},\eta_{k'}^{\dag}\}=0. \quad \quad
\end{eqnarray}
In the Nambu representation, the Schr\"{o}dinger equation $H |\psi_k\rangle=\epsilon_{k}|\psi_k\rangle$ can be recast into a $2N$$\times$$2N$ matrix form as~\cite{Wang2016Phase}
\begin{eqnarray}
 \left(
  \begin{array}{cc}
    A & B \\
    -B^* & -A^T
     \\
  \end{array}
\right)
\left(\begin{array}{c}
    u_{k}\\
    v_{k}
     \\
       \end{array}
     \right)= \epsilon_k \left(\begin{array}{c}
    u_{k}\\
    v_{k}
     \\
       \end{array}
     \right),
     \label{BdGHam}
\end{eqnarray}
where $A=-J(\delta_{i+1,j}+\delta_{i-1,j})+V_i \delta_{i,j}$, $B=-\Delta(\delta_{i+1,j}-\delta_{i-1,j})$, $u_{k}^T = (u_{k,1},\cdots,u_{k,N})$, and $v_{k}^{T}=(v_{k,1},\cdots,v_{k,N})$.
The matrix elements for the boundary terms $A_{N,1}=A_{1,N}=-\sigma J$, $B_{N,1}=-B_{1,N}=-\sigma \Delta$.

The BdG Hamiltonian in Eq.(\ref{BdGHam}) obeys an imposed particle-hole symmetry, implying
$\eta_{k}(\epsilon_k)$=$\eta^\dagger_{k}(-\epsilon_k)$.
The excitation spectrum $\epsilon_{k}$ is a solution of the secular
equation $\mathrm{det}[(A+B)(A-B)-\epsilon_{k}^{2}]=0$.
The energy levels appear in $\pm \epsilon_k$ conjugate pairs, with $\epsilon_{k}\ge0$, except the zero energy mode, which is self-conjugate. In terms of the operators $\eta_{k}$ and $\eta^\dagger_{k}$, the Hamiltonian in Eq.(\ref{Ham}) can be diagonalized as
\begin{eqnarray}
 H=\sum_{k=1}^{N}2\epsilon_{k}(\eta^{\dag}_{k}\eta_{k}-\frac{1}{2}),
\end{eqnarray}
where $\epsilon_{k}$ are single-particle energies in ascending
order, i.e., $\epsilon_{1} \le \epsilon_{2}\le \cdots \le \epsilon_{N}$.
 The ground state of $H$ is the Bogoliubov vacuum state $|\psi_{g}\rangle$
annihilated by all $\eta_{k}$ ($k=1,\cdots,N$), i.e.,  $\eta_{k} |\psi_{g}\rangle$=0, with an energy $E_g$= $-\sum_{k=1}^{N} \epsilon_{k}$.
In the spirit of the Ginzburg-Landau scenario of continuous phase transitions, the characterization of quantum phase transitions
in many-body systems has been traditionally based on a suitable order parameter $Q$, whose expectation value in the ordered phase is finite while it becomes exactly zero at criticality: $Q \sim |V - V_{c}|^{\beta_Q}$, where the exponent $\beta_Q$ of this power law is dubbed as the order parameter critical exponent. The scaling exponents reflect the universality class of the theory, which is independent of the microscopic details of the model but depends only on global properties such as the symmetries and dimensionality of the Hamiltonian. The asymptotic behavior of critical phenomena corresponding to the thermodynamic limit can be retrieved using FSS~\cite{Fisher1972Scaling}:
\begin{eqnarray}
\label{eq:fss0}
\mathcal{Q}(N,V) = N^{-\beta_\mathcal{Q}/\nu}\tilde{\mathcal{Q}}(|V-V_{c}| N^{1/\nu}),
\end{eqnarray}
where $\nu$ characterizes the divergence of the correlation
length and $\tilde{\mathcal{Q}}$ is a universal scaling function~\cite{DeChiara2012Entanglement}.
However, it remains a systematic challenge to identify an explicit local order parameter in random models and analog quasiperiodic systems. There is no explicit symmetry breaking associated with a local order parameter, and therefore has no experimental
signature in the ground-state energy or its derivatives.
Over recent years, an alternative approach to understand
quantum criticalities and the associated universality, exploits the information content stored in the many-body degrees of freedom of a quantum system. In the following, we investigate the quantum criticality of
the AAH model from the quantum information perspective. Quantum entanglement and fidelity susceptibility have been widely exploited in the research of various phase transitions without any prior knowledge of order parameters. Specifically, these information measures can reconcile seemingly unrelated behaviors in different branches of physics ranging from condensed matter physics to gravitational physics. It is interesting to note that quantum entanglement may be used to shed light on the black-hole information paradox regarding
the anti-de Sitter/conformal field theory (AdS/CFT) correspondence, which was initially proposed due to the scaling behavior of entropy in black holes~\cite{Ryu2006Holographic}, while the fidelity susceptibility is dual to the volume of codimension one time slice in AdS spacetime~\cite{Mazhari2017Holographic}.
Therefore, it is quite intriguing to check whether the FSS of the universal order parameters 
applies also in the quasiperiodic models.

\section{VON NEUMANN ENTROPY}\label{sec:vNE}
First attempt along this vein focuses on the study of quantum entanglement. As a pure quantum concept with no classical counterpart, entanglement describes non-local correlations between the constituents of quantum systems.
During the last decade, there has been an increasing interest in the entanglement properties of quantum systems, in particular in specifying quantum criticalities~\cite{Osborne2002Scaling,Vidal2003Entanglement,Gu2004Entanglement}.  
Quantum entanglement is especially suitable for characterizing phases that lack explicit local order parameters, such as topological states~\cite{Hamma2005Bipartite,Kitaev2006Topological}, spin liquid~\cite{Zhang2011Entanglement} and topological order~\cite{Levin2006Detecting}.
It is noted that the outcomes of entanglement witness rely heavily on the possible partition of the Hilbert space, the so-called entanglement cuts, which can be performed on real space, momentum space or internal degrees of freedom of fermions
~\cite{MondragonShemSignatures2014}. The many-body entanglement entropy can be formulated in terms of the covariance matrix restricted to the subsystem, which is the manifestation of the large gaps in the single-particle energy spectrum and very sensitive to the subband structure of the spectrum~\cite{Roy2019Study}.
Despite the extensive literature on the criticality measures related to wavefunction,
there has been only very few work exploring the entanglement entropy in the AAH model
near criticality~\cite{Gong2005Von,Gong2009Localization,Gong2007von,Roosz2014Nonequilibrium}.
The spatial entanglement entropy of a quantum state is a common thread for analyzing the delocalized-localized transitions. In the delocalized phase, where the wave function is extended over many sites, one may
expect considerable correlation spreads in the system and thus the entanglement entropy therein is larger than that in the localized phase. The single-particle states show an interesting multifractal behavior extending to all length scales, which is reflected by the nonlinear dependence of the fractal dimensions $D_q$ on $q$, $q \ge 0$, defined via
the scaling of the $q$-order IPR:
\begin{eqnarray}
\label{Dq}
D_q=\left\langle \frac{\ln \sum_j \vert\psi_{k,j} \vert^{2q}}{(1-q) \ln N}\right\rangle,
\end{eqnarray}
where $\langle \cdot \rangle $ denotes the average over all $k$-th eigenstates. By performing the FSS of the mean IPR, the fractal dimensions $D_{2}= 1$ and 0 characterize the self-similar behavior for the EP and the LP, respectively, whereas  $0< D_{2} < 1$ implies multifractality in critical states. We are aware of the fact that the entanglement entropy has not yet been employed to analyze the multifractal structure in the AAH model with $p$-wave pairing, 
although an upper bound for the entanglement entropy for any eigenstate with a given fractal dimension was established~\cite{DeTomasi2020Multifractality}.

To this end, we here conduct the analysis of the single-site entanglement entropy.
For spinless noninteracting fermions in
the AAH model, there are two local states at the $j$-th site, i.e., $|1\rangle_{j}$, $|0\rangle_{j}$, corresponding to states with one and zero particles, respectively.
Considering a generic eigenstate for Hamiltonian (\ref{Ham})
that can be obtained by diagonalizing Eq.(\ref{BdGHam}),
the single-site reduced density matrix $\rho_{k,j}$ for $k$-th eigenstate at the $j$-th site can be written as
 \begin{eqnarray}
\rho_{k,j}=|u_{k,j}|^{2}|1\rangle_{j} \langle1|_{j}+(1-|u_{k,j}|^{2})|0\rangle_{j} \langle 0|_{j}.
\label{densitymatrix}
\end{eqnarray}
Consequently, the single-site von Neumann entropy associated with site $j$
can be expressed as
\begin{eqnarray}
S_{k,j}&=& -|u_{k,j}|^{2}\ln  |u_{k,j}|^{2}
- (1-|u_{k,j}|^{2}) \ln (1-|u_{k,j}|^{2}. \label{von Neumann entropy} \quad
\end{eqnarray}
For quasiperiodic system, the spatial von Neumann entropy for $k$-th eigenstate over different sites is given by
\begin{eqnarray}
S_{k}=\sum_{j=1}^{N} S_{k,j},
\label{site-averaged von Neumann entropy}
\end{eqnarray}
and the spectrum averaged entanglement entropy (SAEE)
\begin{eqnarray}
\langle S \rangle=\frac{1}{2N}\sum_{k=1}^{2N} S_{k}.
\label{averaged Neumann entropy}
\end{eqnarray}

\begin{figure}[tb]
\centering
\includegraphics[width=\columnwidth]{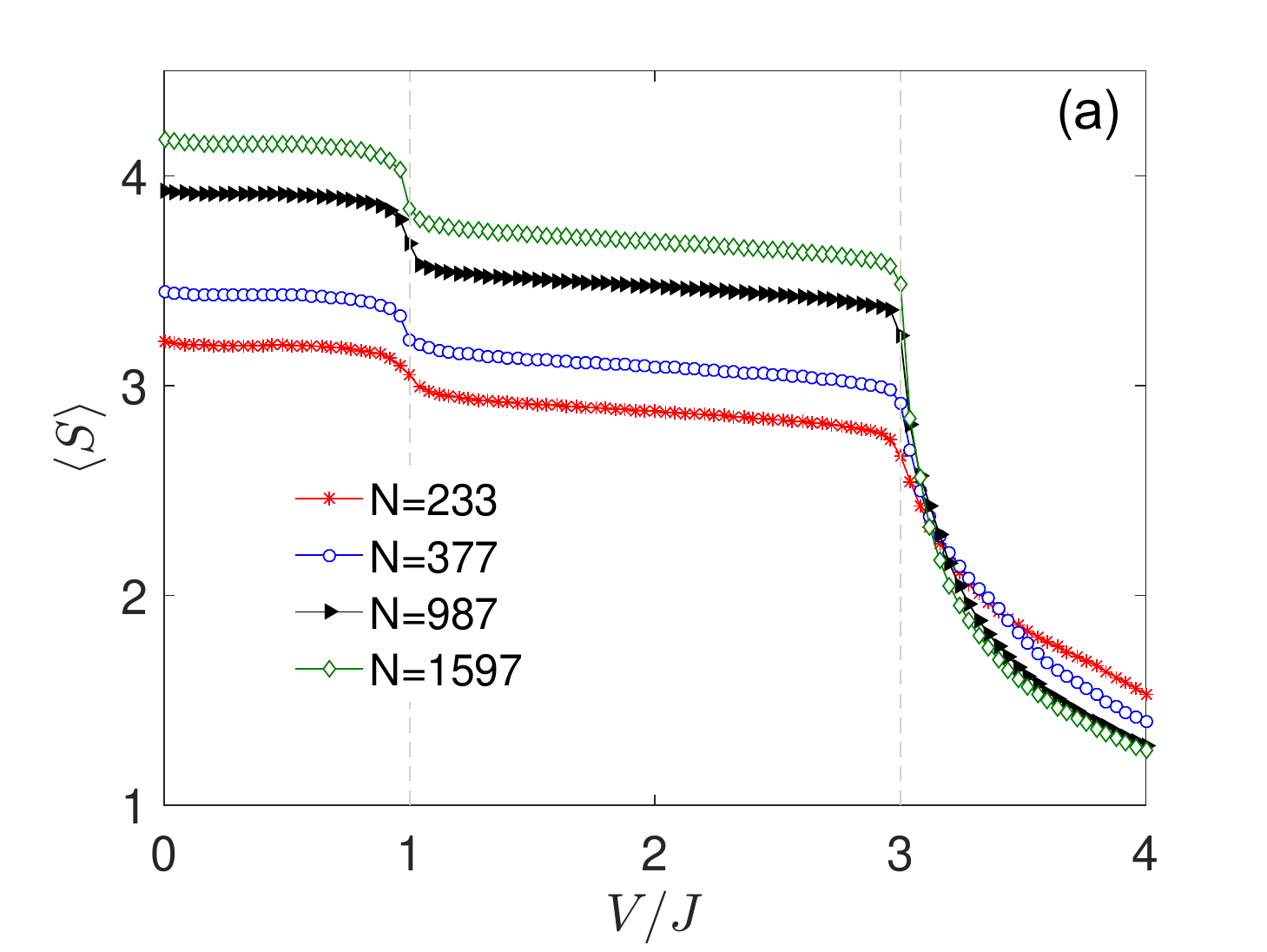}
\includegraphics[width=\columnwidth]{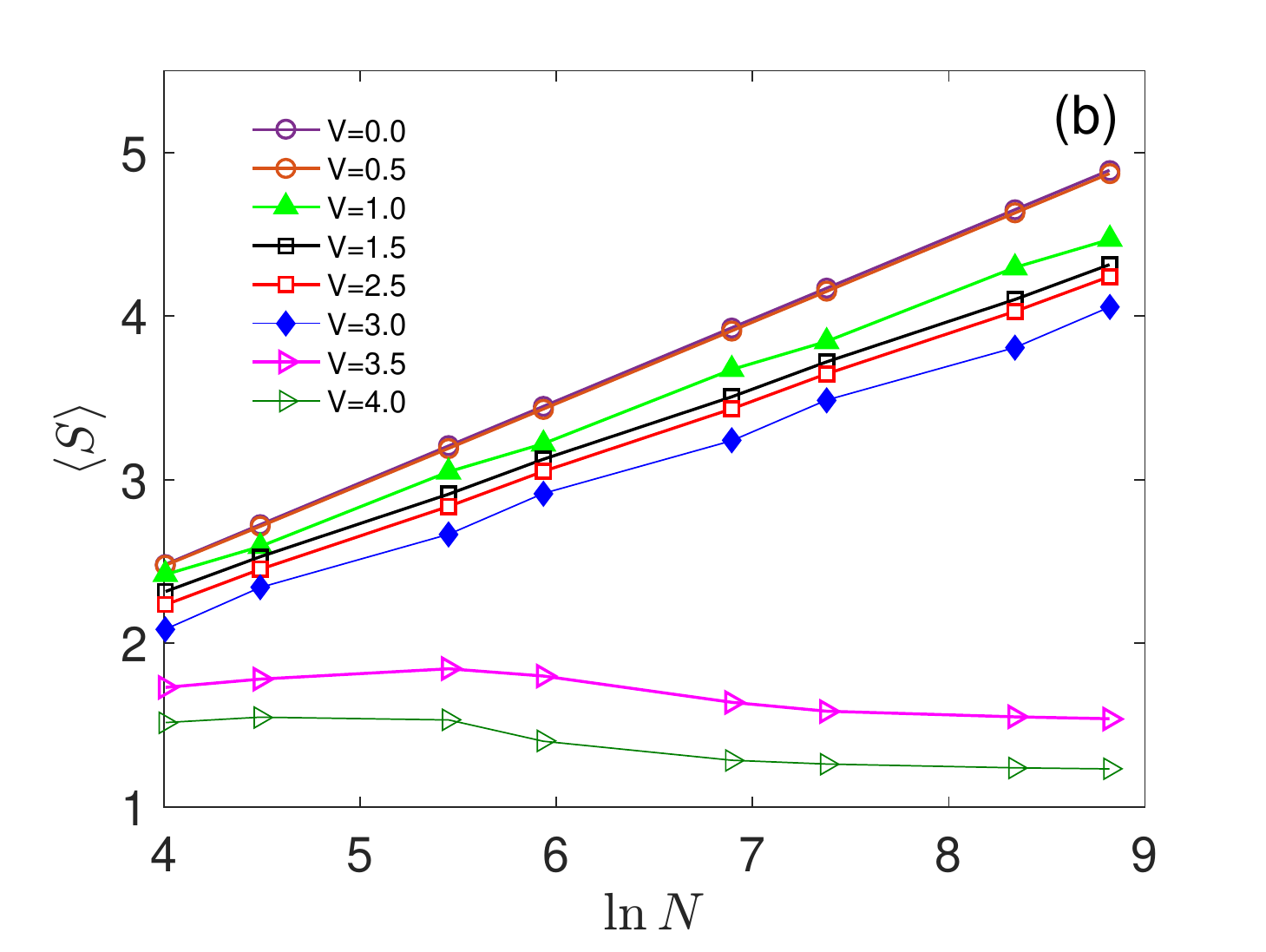}
 \caption{(a) The SAEE $\langle S \rangle$ versus the incommensurate potential strength $V$ for various system sizes $N$.  (b) The scaling of $\langle S \rangle$ for different $V$. Here periodic boundary conditions are used with $\Delta=0.5$, $\phi=0$. }
  \label{fig:AAHaveragedS}
\end{figure}

Figure \ref{fig:AAHaveragedS}(a) shows the SAEE as a function of the incommensurate potential $V$ for various system sizes $N$ with $\Delta = 0.5$ and $\phi=0$.
One can observe $\langle S \rangle$ displays sudden falls
at $V_{c1} = 2|J-\Delta |$ and $V_{c2} = 2|\Delta + J|$.
It is clear that $\langle S \rangle$ has a lower value in the LP than that in the delocalized phases. In the EP and the CP, $\langle S \rangle$ shows a monotonic increase as the system sizes $N$ increase, in stark contrast to the decreasing tendency in the LP. We then assume an ansatz of the FSS for $\langle S \rangle$ in the vicinity of $V_{c1}$ under investigation~\cite{Fisher1972Scaling}:
\begin{eqnarray}
\label{eq:fss}
\langle S \rangle = N^{\beta_S/\nu}\tilde{S}(|V-V_{c1}| N^{1/\nu}),
\end{eqnarray}
where $\beta_S$ is the corresponding critical exponent,  and $\tilde{S}$ is a universal scaling function~\cite{DeChiara2012Entanglement}.

\begin{figure}[tb]
\centering
\includegraphics[width=\columnwidth]{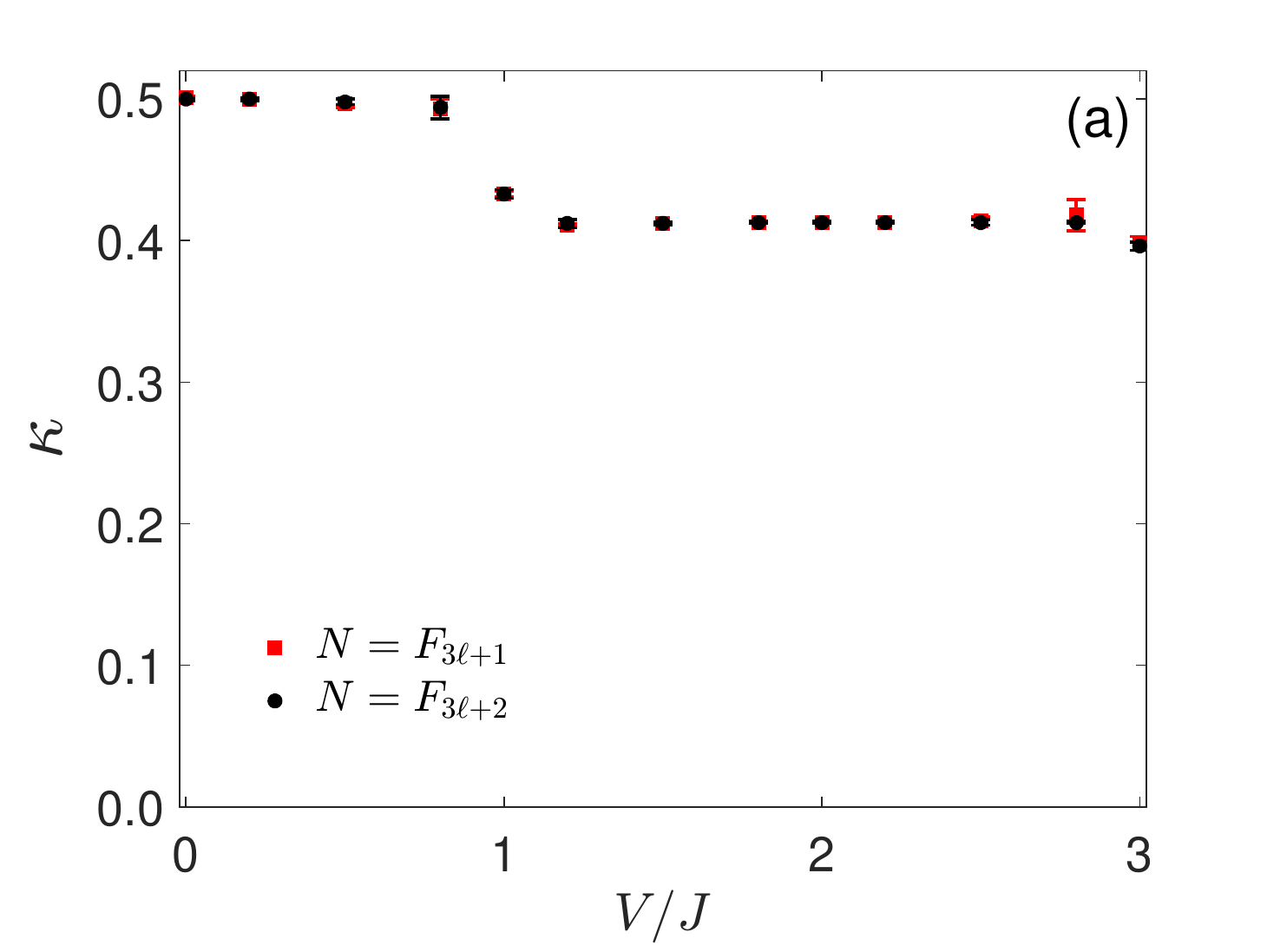}
 \includegraphics[width=\columnwidth]{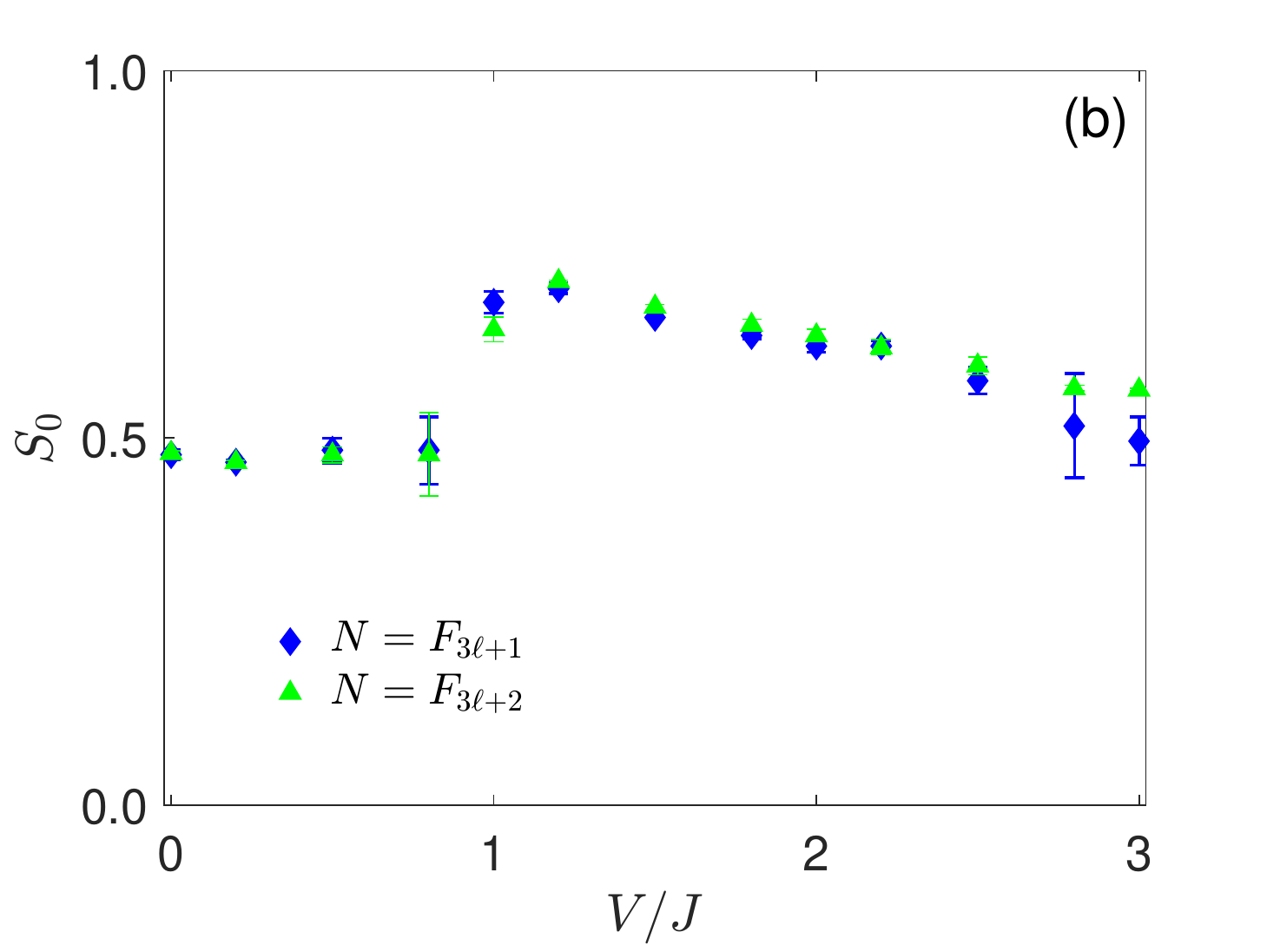}
 \caption{The fitted values  of (a) $\kappa$ and (b) $S_0$ in  Eq.(\ref{scaling_averaged Neumann entropy}) for $N=F_{3\ell+1}$ and $F_{3\ell+2}$. Here periodic boundary conditions are used with $\Delta=0.5$, $\phi=0$.}
  \label{fig:kappa}
\end{figure}

Figure \ref{fig:AAHaveragedS}(b) reveals that the SAEE in the delocalized phases follows the law
\begin{eqnarray}
\langle S \rangle=\kappa \ln N + S_0,
\label{scaling_averaged Neumann entropy}
\end{eqnarray}
with the asymptotic prefactor $\kappa$ and the nonuniversal residual entropy $S_0$. The fitted values $\kappa$ and $S_0$ are displayed in Fig.\ref{fig:kappa} for different quasiperiodic potential strengths. For $V<V_{c1}$, $\langle S \rangle$ has a negligible dependence on $V$. A close inspection for $N$ up to $6765$ finds $\kappa_{\rm EP} \approx 0.500$, which is almost identical with the
mean spectral exponent $\alpha_0=0.5$~\cite{Evangelou2000Critical}.
Especially for the homogeneous case $V=0$, the system is simply a single-band Kitaev Hamiltonian.
The spatial entanglement entropy becomes equivalent to Meyer-Wallach measure of multipartite global entanglement~\cite{Montakhab2010Multipartite,Radgohar2018Global}.
While for $V_{c1}<V<V_{c2}$, the overall logarithmic scaling of $\langle S \rangle$ is decorated with an oscillation of a two-subsequence period. One has to fit the data for $N=F_{3\ell+1}$ and $N=F_{3\ell+2}$ separately. 
Despite wide fluctuations, the fitted slope is nearly a constant as  $\kappa_{\rm CP} \approx 0.414$.  In fact, in the limit $q \to 1$, Eq.(\ref{Dq}) provides important information on
the effective dimension of the support set for the mean entanglement entropy in Eq.(\ref{averaged Neumann entropy}), which scales as $\langle S \rangle \sim N^{D_1}$ given by ${D_1}=\kappa_{\rm CP}/\kappa_{\rm EP}$ in the
Nambu space~\cite{Siebesma1987Multifractal}. We note that the value is close to the maximal fractal dimension $D_2 \approx 0.82$ in the ground
state of the Harper model~\cite{Evangelou2000Critical}.
The fluctuations are reflected in the fitted values of $S_0$, which  varies with $V$. One observes $\langle S \rangle$ first grows and then declines until saturating to finite values in the limit $N \to \infty$, thus yielding $\kappa_{\rm LP} \to 0$ in Eq.(\ref{scaling_averaged Neumann entropy}) for localized
wave functions. 
We can anticipate $\langle S \rangle  \approx 0$ in the extremely
localized phase for $V \to \infty$.
In the localized phase, all one-particle eigenstates are localized on a length $N_{\rm loc}$. For lengths larger than the localization length $N_{\rm loc}$, the wave function consists of a single isolated structure and its correlation dimension $D_2$ is zero. One may expect that in large systems $N \gg N_{\rm loc}$ the length scale is set by the localization length $N_{\rm loc}$ rather than the system size $N$, the entanglement entropy in Eq.(\ref{scaling_averaged Neumann entropy}) may be rewritten as $\langle S \rangle=\kappa_{\rm loc} \ln N_{\rm loc} + S_0$~\cite{Roosz2020Entanglemen}. The ansatz is expected to deteriorate when the diverging localization length becomes comparable with the system size.

\begin{figure}[tb]
\centering
\includegraphics[width=\columnwidth]{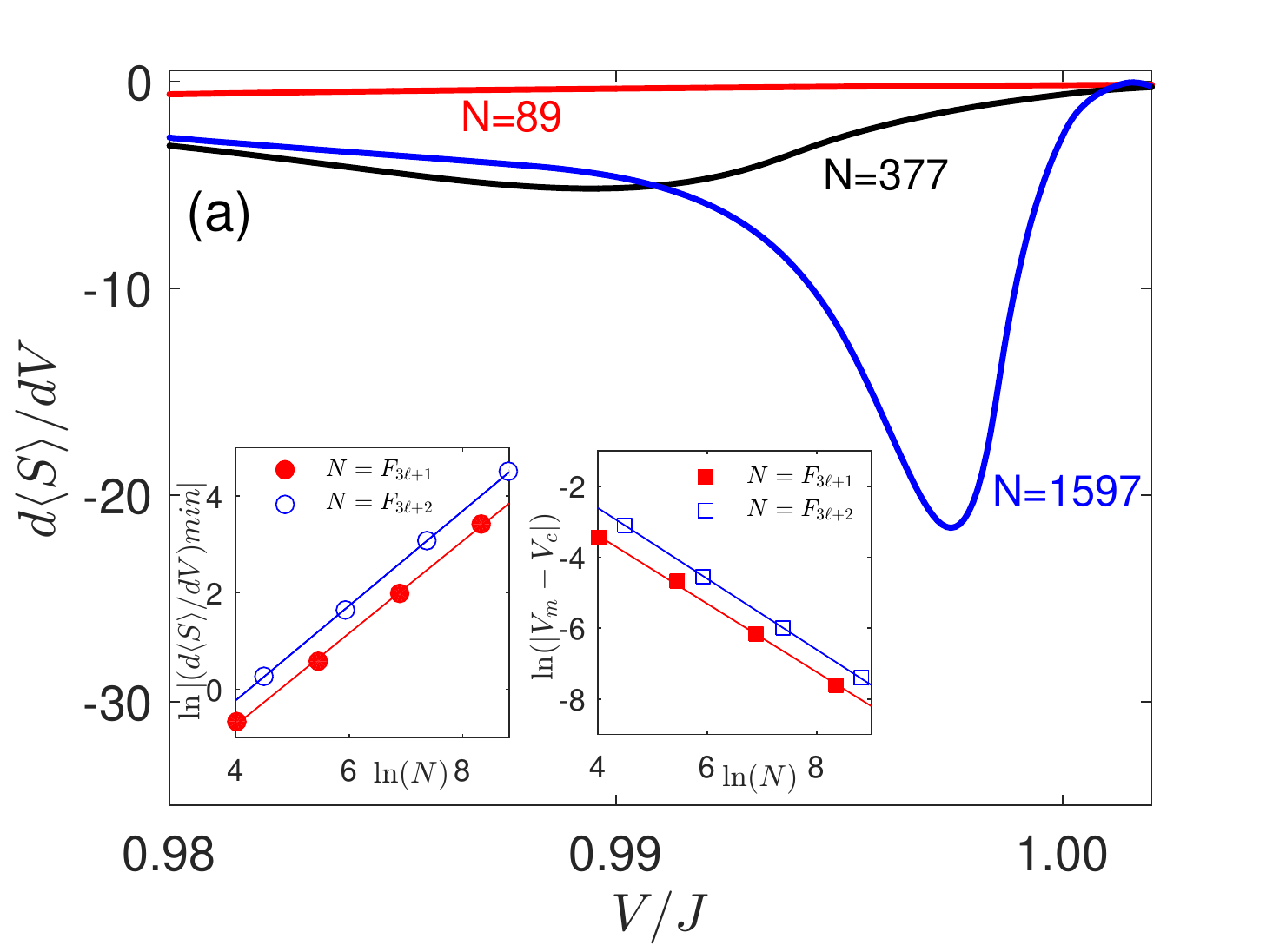}
 \includegraphics[width=\columnwidth]{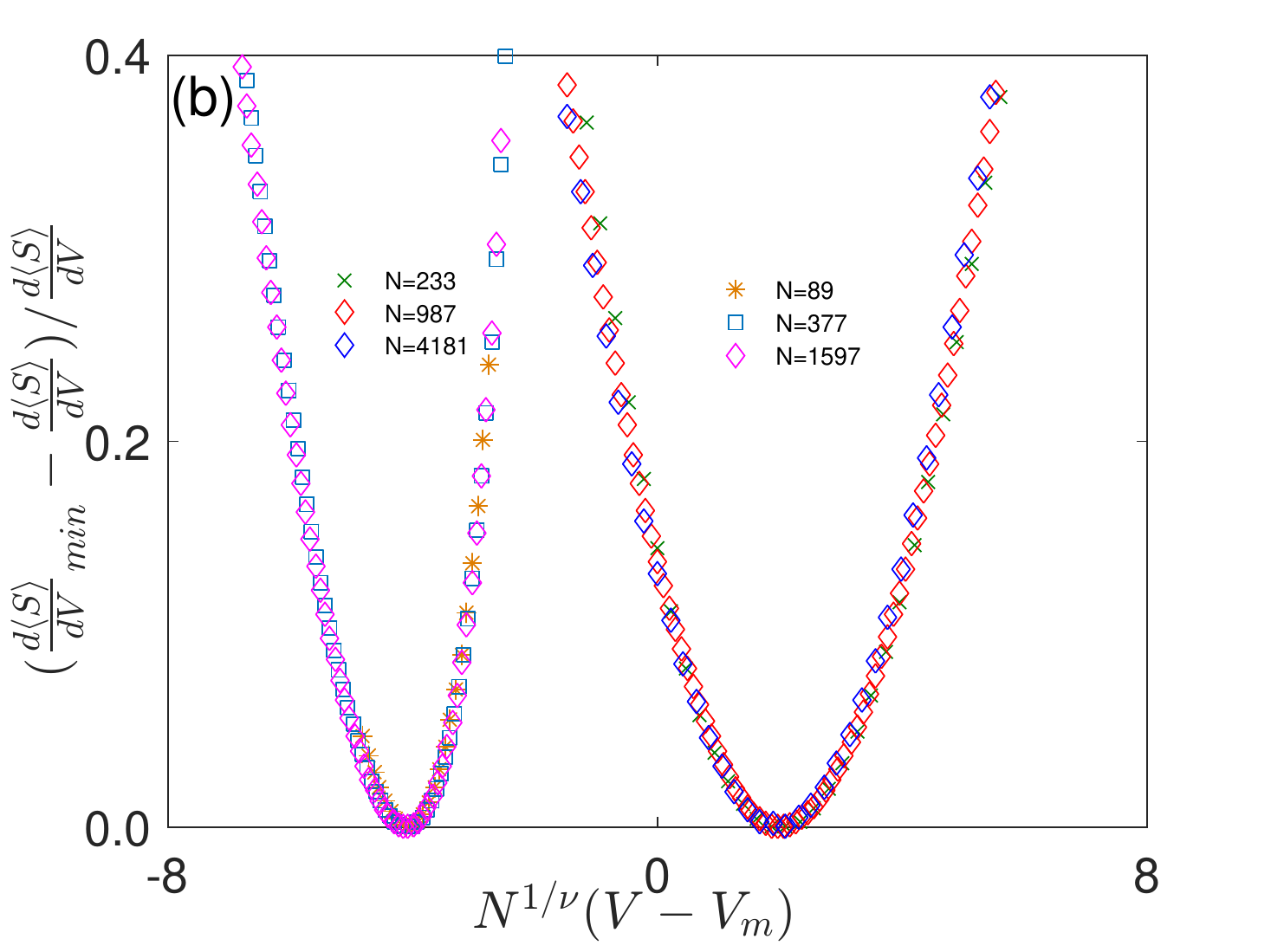}
 \caption{(a) $d\langle S \rangle/dV$ with respect to $V$ with the system sizes $N=89, 377, 1597$. The left and right insets show  the scaling behavior of $\ln \vert (d \langle S \rangle/dV)_{\rm min} \vert$ and  $\ln(|V_{m}-V_{c}|)$ for $N$.
 (b) $\left[\left(d\langle S \rangle/dV\right)_{\rm min}-\left(d \langle S \rangle/dV\right) \right]$/$\left(d\langle S \rangle/dV\right)$  as a function of the scaled variable $N^{1/\nu}(V-V_{m})$. All curves for odd number of lattice sizes collapse into two separate curves when we choose the correlation-length critical exponent $\nu=1.000$. Here  $\Delta=0.5$ and $\phi=0$. }
  \label{fig:dSdV}
\end{figure}
It is evident that Eq.(\ref{scaling_averaged Neumann entropy}) implies that $\beta_S=0$ in the vicinity of the EP-CP transition.
To refine more critical exponents, we have to resort to the first derivative of the SAEE as
 \begin{eqnarray}
\label{eq:fss3}
d\langle S \rangle/dV = N^{(1+\beta_S)/\nu}\tilde{S}^{'}(|V-V_{c1}| N^{1/\nu}),
\end{eqnarray}
where $\tilde{S}^{'}(\cdot)$ is the first derivative with respect to the argument.
Using the ansatz in Eq.(\ref{eq:fss3}),  we perform the FSS for odd number of system sizes around the critical point $V_{c1}$. As shown in Fig.\ref{fig:dSdV}(a), the minima of $\mathrm{d} \langle S \rangle / \mathrm{d} V$ become deeper and the corresponding positions of the valleys $V_{m}$ converge towards the critical points with increasing the system sizes. A careful analysis easily identifies $V_{m}$ are above $V_{c1}$ for $N=F_{3\ell+1}$,  while they are below $V_{c1}$ for $N=F_{3\ell+2}$. The two-subsequence behavior of $d\langle S \rangle/dV$ can be manifested by taking logarithm on both sides of the ansatz (\ref{eq:fss3}),
\begin{eqnarray}
\label{eq:fss1}
\ln d\langle S \rangle/dV  = [(1+\beta_S)/\nu] \ln N+\ln \tilde{S}^{'}(|V-V_{c1}|N^{1/\nu}).\quad \quad
\end{eqnarray}
To be concrete, the power-law relations can be further identified as
 \begin{eqnarray}
\label{eq:fss4}
\left\vert \left(\frac{\mathrm{d} \langle S \rangle }{ \mathrm{d} V}\right)_{\rm min}\right\vert&=& \tilde{S}^{'}(0) N^{(1+\beta_S)/\nu},
 \\
|V_m-V_{c1}|&\propto& N^{-1/\nu}.\label{eq:fss5}
\end{eqnarray}
The linear scaling relations
are confirmed by the numerical fittings in the insets of Fig.\ref{fig:dSdV}(a).
The extrema of $d\langle S \rangle/dV$ satisfy
$\ln(\mathrm{d} \langle S \rangle / \mathrm{d} V)_{\rm min}$ = $(0.953 \pm 0.095)$  $\ln N$-$(4.552\pm 0.608)$ and $\ln(  V_m - V_{c1})$=$(-0.962\pm 0.117) \ln N +(0.458\pm 0.169)$ for $N=F_{3\ell+1}$, suggesting $\beta_S=0$, $\nu=1.000$.
Similarly, for $N=F_{3\ell+2}$ the linear fit to the log-log plot yields $\ln(\mathrm{d} \langle S \rangle / \mathrm{d} V)_{\rm min} $ = $(0.980 \pm 0.027)$ $\ln N$-$(4.147\pm 0.183)$ and $\ln(V_{c1} -V_m )$=$(-0.998\pm 0.010) \ln N +(1.378\pm 0.071)$. Apparently the
 estimated value of critical exponents agree well with each other, which are independent of Fibonacci subsequence. Figure \ref{fig:dSdV}(b) shows that the scaled $d\langle S \rangle/dV$ near criticality for different values of $N$ superposes onto two separate scaling functions when the correlation-length critical exponent $\nu = 1.000$ is chosen. The perfect data collapse validates the scaling relation in Eq.(\ref{eq:fss3}). 

\section{Generalized Fidelity Susceptibility}\label{sec:GFS}
It has been shown that the multipartite entanglement dictates not only the position of critical points but also
the correlation-length critical exponent $\nu$. However, according to Widom scaling hypothesis~\cite{Widom1965Equation}, there are generally two independent critical exponents and thus a second independent critical exponent plays a decisive role in determining the universality class. We then endeavor to apply GFS to obtain the dynamical exponent $z$, which has been successfully developed in the CP-LP transition of the $p$-wave-paired AAH model~\cite{Lv2022Quantum}.
The fidelity susceptibility has been regarded as a vital tool to identify critical points by measuring the rate of change
for a given state
after a sudden infinitesimal quench of the tuning parameter~\cite{You2007Fidelity}. It should be emphasized that the ground-state fidelity susceptibility is unable to witness the EP-CP transition, as is shown in the Appendix \ref{sec:GSFS}.  As a result, the GFS associated with an eigenstate $\vert \psi_k\rangle$
has a visible implication on the response of the system,
which is given by a summation form~\cite{You2015Generalized,GU2010FIDELITY}
 \begin{eqnarray}
 \chi_{2r+2}^{(k)} = \sum_{k'\neq k} \frac{\vert \langle  \psi_{k'} \vert \partial_V \hat{H}  \vert \psi_k   \rangle \vert^2}{(\epsilon_{k'}  -\epsilon_k )^{2r+2}},
\label{generalizedsusceptibility}
\end{eqnarray}
where $\vert \psi_{k'}\rangle $ and $\epsilon_{k'}$ correspond to the $k'$-th eigenstate and eigenvalue of Hamiltonian (\ref{Ham}), respectively. In parallel with Eq.(\ref{averaged Neumann entropy}), it is tempting to speculate that the spectrum averaged fidelity susceptibility serves as an effective tool for characterizing the quantum criticality in the AAH model. In Appendix \ref{sec:SAFS}, we unveil that there is no qualitative difference between the average fidelity susceptibility and the typical fidelity susceptibility.

In this following, we focus on the GFS of the lowest eigenstate $\vert\psi_1\rangle $. In this case, Eq.(\ref{generalizedsusceptibility}) reduces respectively to the second derivative of $\chi_1\equiv \partial^2 \epsilon_1/\partial V^2$ for $r=-1/2$  and the standard fidelity susceptibility $\chi_2\equiv \langle \partial_V \psi_1 \vert \partial_V \psi_1 \rangle - \langle \partial_V \psi_1 \vert  \psi_1\rangle \langle \psi_1 \vert  \partial_V \psi_1 \rangle$
for $r=0$~\cite{You2007Fidelity}.
Accordingly, the GFS of a finite system with size $N$ in the neighborhood of $V_{c1}$ shall obey the universal scaling form~\cite{Quantum2010Albuquerque}:
\begin{eqnarray}
\chi_{2r+2}=N^{\beta_{r}}\tilde{\chi}_{r}(\vert V-V_{c1}\vert N^{1/\nu}),
\label{eq:chi_scaling}
\end{eqnarray}
where $\beta_{r}\equiv2/\nu+2zr$ is the critical adiabatic dimension, and $\tilde{\chi}_{r}$ is a regular universal scaling function of the GFS of order $2r+2$. With increasing the system sizes $N$, the peaks of GFS become sharper and the peak position $V_{m}$ approaches the critical point.
However, in the actual calculation, $V_m$ is quite close to $V_{c1}$ for a moderate $N$ owing the accelerated convergence of the fidelity susceptibility and thus Eq.(\ref{eq:fss5}) is beyond the current numerical accuracy. The determination of critical exponents can be only recapitulated through the following relation:
\begin{eqnarray}
\chi_{2r+2}(V_{m})&=& \tilde{\chi}_{r}(0) N^{\beta_{r}}. \label{eq:chi_max}
\end{eqnarray}
\begin{figure}[tb]
\centering
\includegraphics[width=\columnwidth]{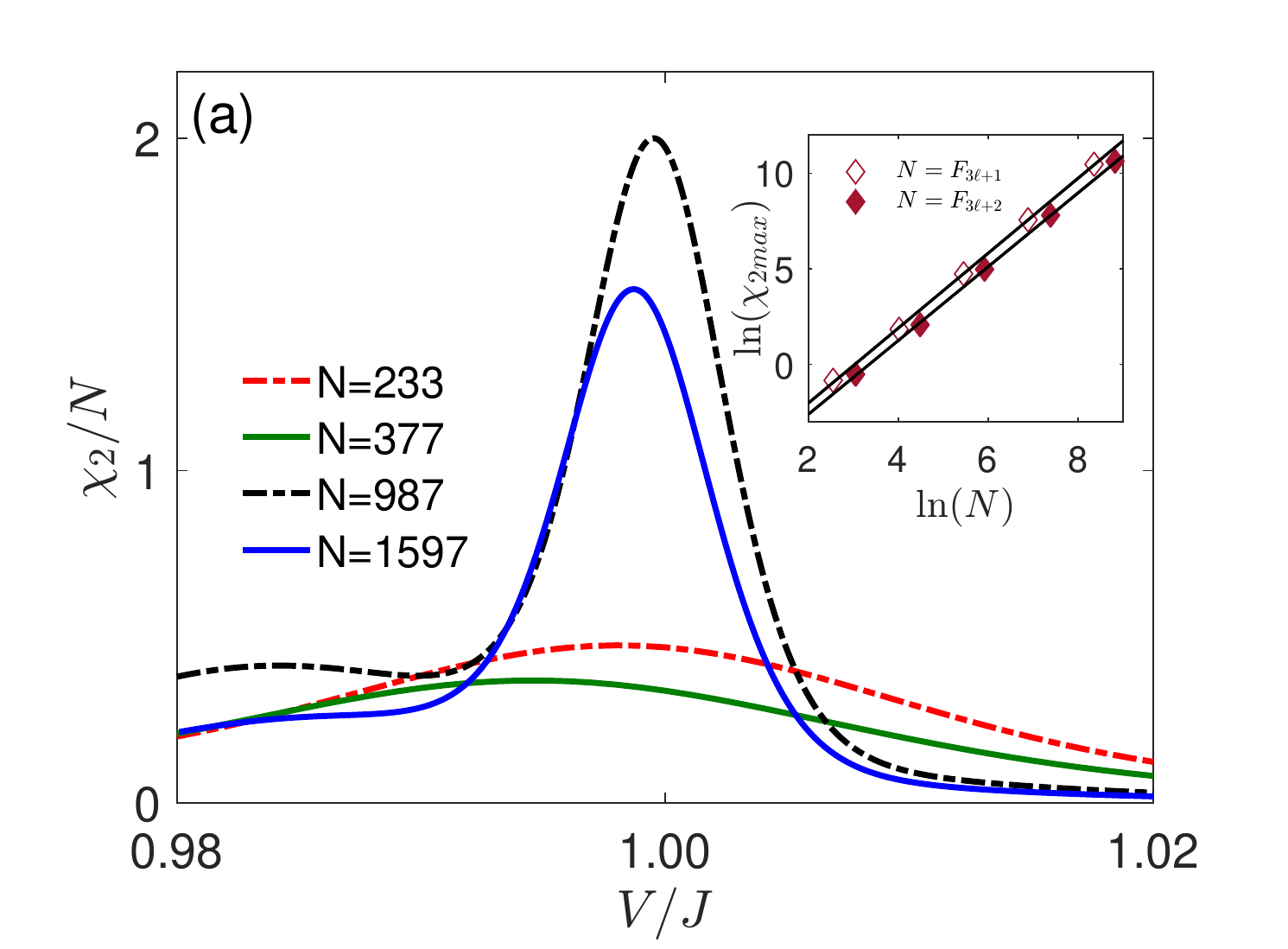}
\includegraphics[width=\columnwidth]{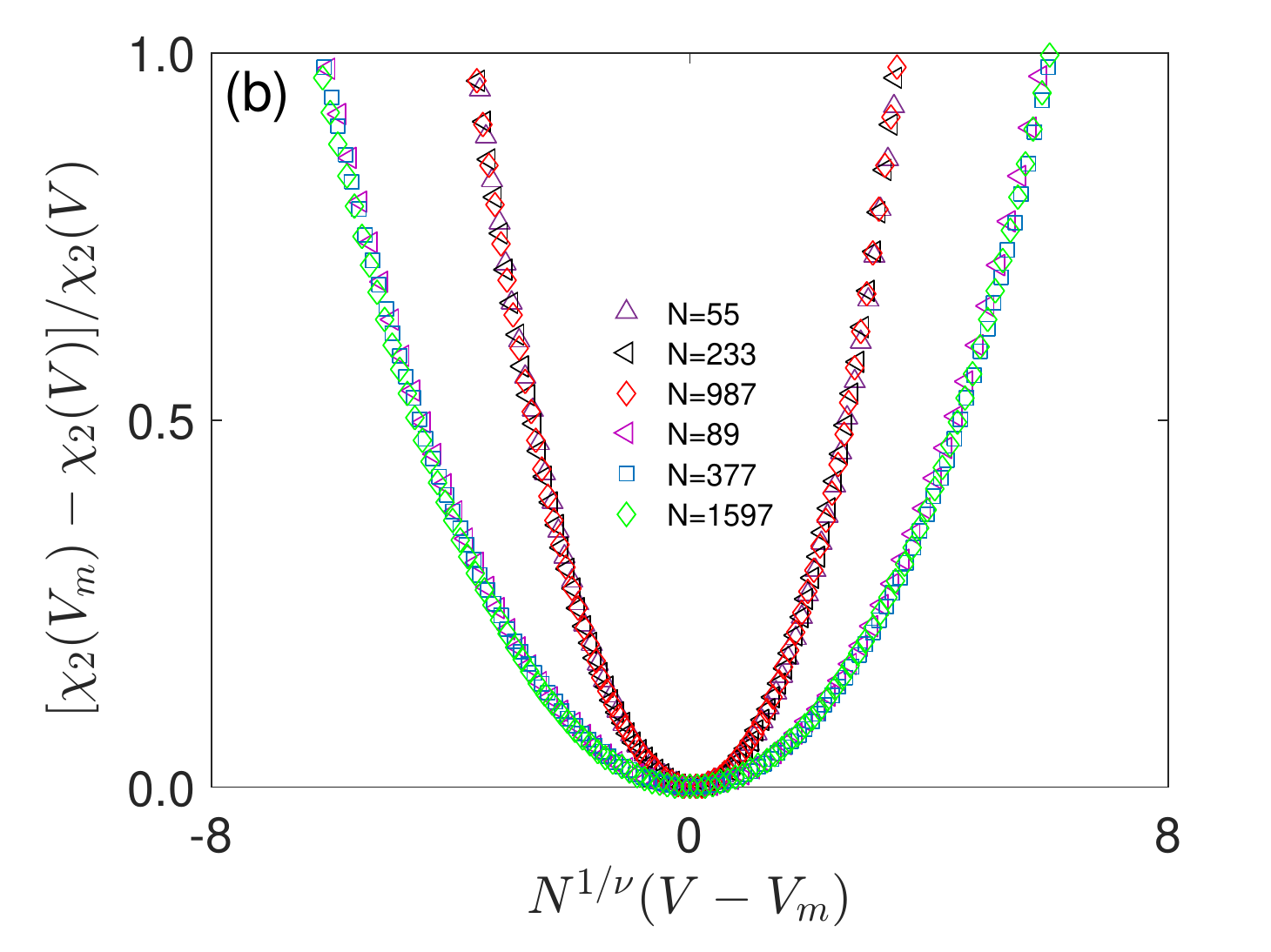}
 \caption{(a) The fidelity susceptibility per site $\chi_{2}/N$ as a function of the incommensurate potential strength $V$ around $V_{c1}=1.0$. The inset shows the scaling behavior of the maxima versus the system sizes for $N=F_{3\ell+1}$: $13$, $55$, $233$, $987$, $4181$ and $N=F_{3\ell+2}$: $21$, $89$, $377$, $1597$, $6765$. (b) Scaled fidelity susceptibility $[\chi_{2}(V_m)-\chi_{2}(V)]/\chi_{2}(V)$ as a function of the scaled variable $N^{1/\nu}(V-V_{m})$. All curves collapse into two separate curves when we choose the correlation-length critical exponent  $\nu=1.000$. Here periodic boundary conditions are used with $\Delta=0.5$, $\phi=0$.}
  \label{fig:AAHmodelfs2}
\end{figure}

Considering the fidelity susceptibility is proportional to the system size far away from the critical point, we show the fidelity susceptibility per site $\chi_{2}/N$ as a function of the incommensurate potential $V$ for $\Delta=0.5$ in Fig.\ref{fig:AAHmodelfs2}(a). The peaks of the fidelity susceptibility around the critical point $V_{c1}=1.0$ become more pronounced for increasing the system sizes $N$. The maxima of $\chi_{2}$ show a power-law divergence, manifested by a linear fit between the maximum of $\ln\chi_{2}$ and $\ln N$. We can also observe a two-subsequence behavior and the fitted slopes give rise to $\beta_0\equiv 2/\nu =1.991 \pm 0.007$ ($1.972 \pm 0.035$) for $N=F_{3\ell+1}$ ($F_{3\ell+2}$) according to  Eq.(\ref{eq:chi_max}). Therefore, the retrieved correlation-length exponent corresponds to $\nu=1.005\pm 0.018$ ($1.014\pm 0.018$) for $N=F_{3\ell+1}$  ($F_{3\ell+2}$). To coin the single-parameter scaling hypothesis in Eq.(\ref{eq:chi_scaling}), we also plot the rescaled fidelity susceptibility $[\chi_{2}(V_{m})$-$\chi_{2}(V)]/\chi_{2}(V)$ as a function scaled variable $N^{1/\nu}(V-V_{m})$. When $\nu=1.000$ is chosen, the curves with distinct system sizes in the vicinity of $V_{c1}$ collapse onto two scaling functions, as are evinced in Fig.\ref{fig:AAHmodelfs2}(b).

\begin{figure}[tb]
\centering
\includegraphics[width=1.\columnwidth]{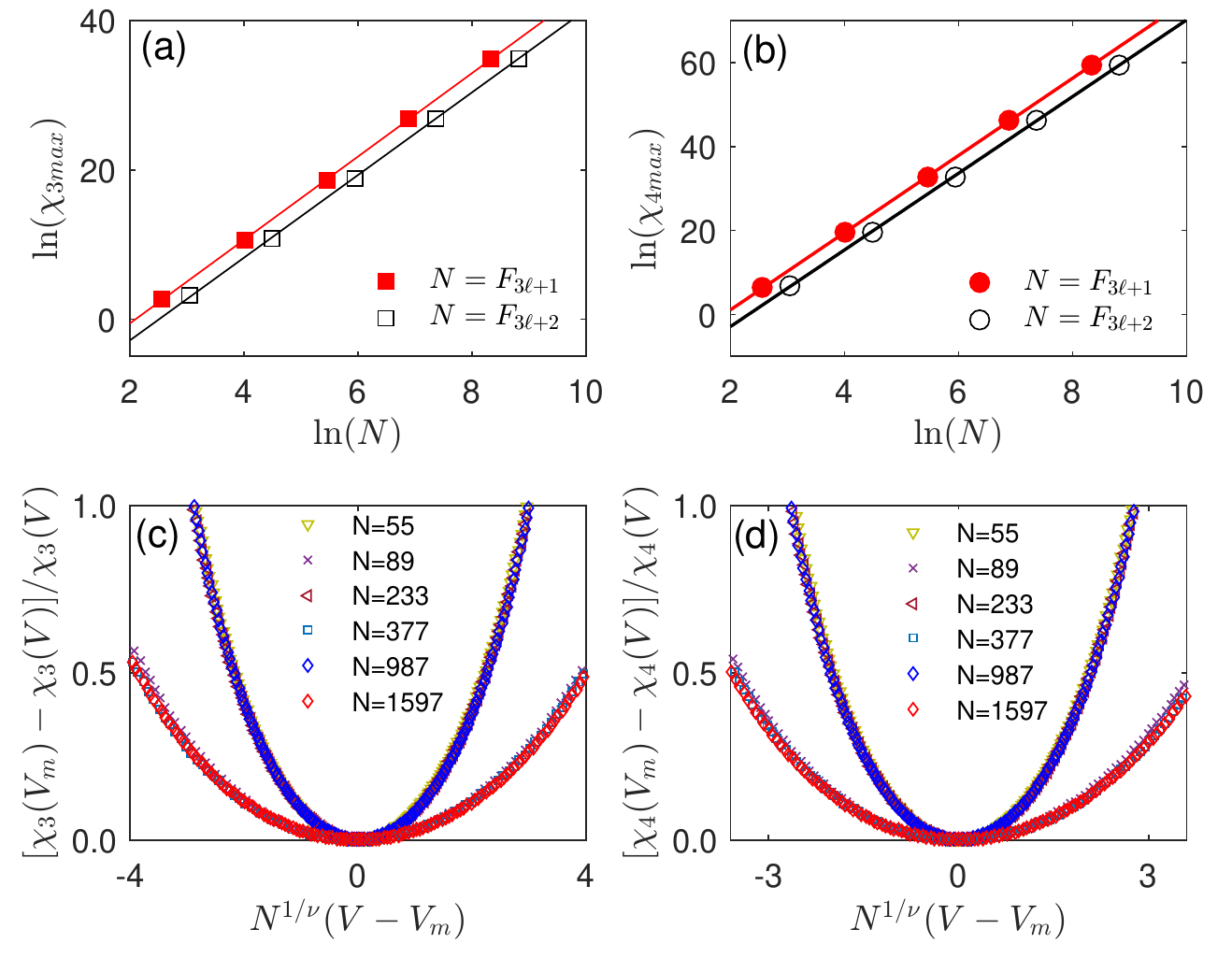}
\caption{ Scalings of (a) $\chi_{3,{\rm max}}$ and (b) $\chi_{4, {\rm max}}$  with respect to system sizes $N$.
Scaled fidelity susceptibilities (c) $[\chi_{3}(V_m)-\chi_{3}(V)]/\chi_{3}(V)$ and (d) $[\chi_{4}(V_m)-\chi_{4}(V)]/\chi_{4}(V)$ as a function of the scaled variable $N^{1/\nu}(V-V_{m})$ for odd number of lattice sizes. All curves are collapsed into two separate scaling functions: one for $N=F_{3\ell+1}$ and the other for $N=F_{3\ell+2}$ when we choose the correlation-length critical exponent $\nu=1.000$.  Here we take periodic boundary conditions with $\Delta=0.5$ and $\phi=0$. }
\label{fig:GFS}
\end{figure}
\begin{figure}[tb]
\centering
\includegraphics[width=\columnwidth]{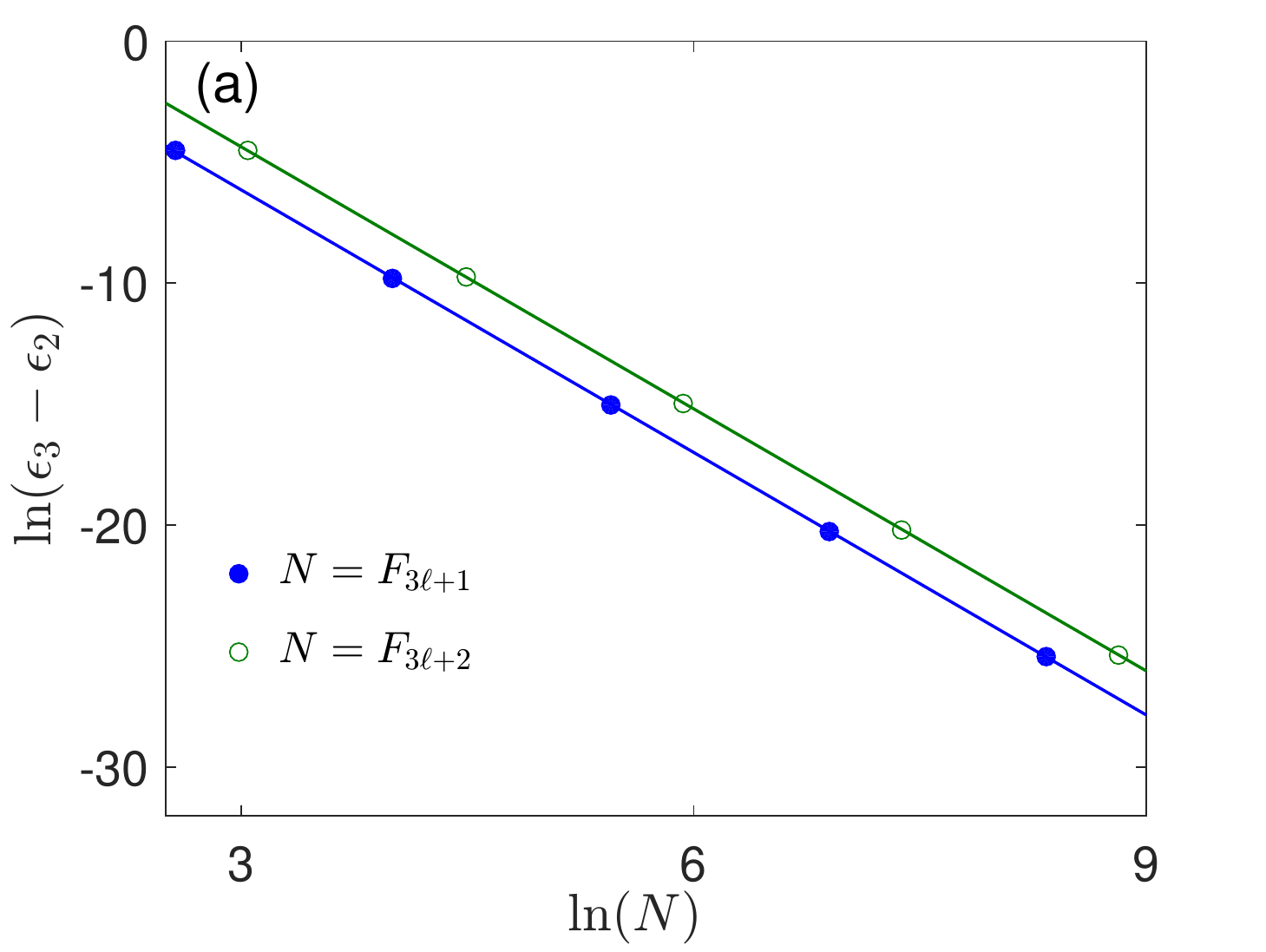}
\includegraphics[width=\columnwidth]{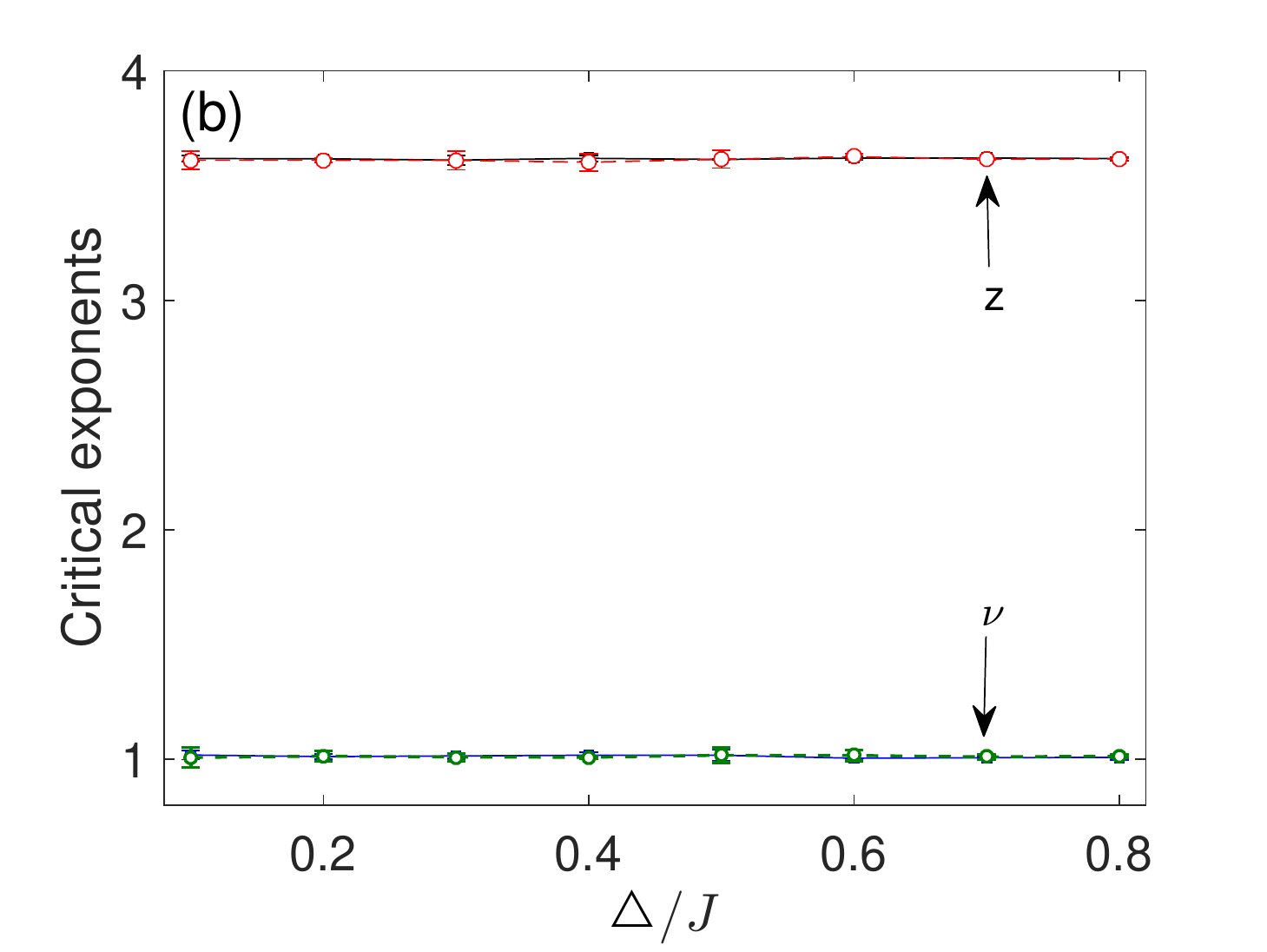}
 \caption{ (a)  The scaling behavior of the
minima of $\epsilon_r \equiv\epsilon_{3}-\epsilon_{2}$ versus $N$  around the critical point $V_{c1}=1.0$ for $\Delta=0.5$.  (b) The fitted values of critical exponents $\nu$ and $z$ as $\Delta$ varies for $N=F_{3\ell+1}$ (solid symbols) and $F_{3\ell+2}$ (open symbols). Here we take periodic boundary conditions with $\phi=0$.}
  \label{fig:AAHmodelerrorbar}
\end{figure}

In order to extract the dynamical exponent $z$ of the transition between the EP and the CP, we further study higher-order GFSs.
One can easily notice that $\chi_{3}$ and $\chi_{4}$  display much more divergent peaks than $\chi_{2}$ in the vicinity of the critical point, showing that the higher-order GFSs
are more efficient in spotlighting the pseudocritical points. The scaling behaviors between $\ln \chi_{3,{\rm max}}$ and $\ln N$ are displayed in Fig.\ref{fig:GFS}(a).
According to Eq.(\ref{eq:chi_max}), the linear fittings of the log-log plot give rise to $\beta_{1/2}\equiv 2/\nu +z=5.607 \pm 0.009$ $(5.607 \pm 0.009) $ for $N=F_{3\ell+1}$ ($F_{3\ell+2}$), corresponding to the dynamical exponent $z \equiv \beta_{1/2}-\beta_{0}=3.616\pm 0.002$ ($3.615\pm 0.018$). Subsequently, the fitting lines of $\ln \chi_{4,{\rm max}}$ with respect to $\ln N$ are exhibited in Fig.\ref{fig:GFS}(b), whose slopes
yield $\beta_{1}\equiv 2/\nu +2z=9.220 \pm 0.007$ ($9.201 \pm 0.023$) for $N=F_{3\ell+1}$ ($F_{3\ell+2}$), corresponding to $z\equiv (\beta_{1}-\beta_{0})/2 =3.615\pm 0.001$ ($3.615\pm 0.006$).
The curves for $\chi_{3}$ and $\chi_{4}$
near the EP-CP transition can be separately rescaled onto two different universal curves for odd numbers of lattice sites with the same exponent $\nu=1.000$, as seen in Figs.\ref{fig:GFS}(c) and (d).
Since there is no gap scaling of $\epsilon_1$ around $V_{c1}$, a relevant spectral gap can be defined as $\epsilon_r \equiv\epsilon_{3}-\epsilon_{2}$.
Fitting $\epsilon_r$ with respect to the system size $N$ yields $z=3.610 \pm 0.017$ ($-3.617 \pm 0.022$) for $N=F_{3\ell+1}$ ($F_{3\ell+2}$), as is revealed in Fig.\ref{fig:AAHmodelerrorbar}(a).
As such, the extracted value of the dynamical exponent $z$ is in perfect agreement with the GFS scaling.

Subsequently, we proceeded to extract $\nu$ and $z$ for different $\Delta$ utilizing the same strategy in Fig.\ref{fig:AAHmodelerrorbar}(b). It is found that the critical-length exponent $\nu \approx 1.000$ and the dynamical exponent $z \approx 3.610$ are almost unchanged for varying $\Delta$. We note that the difference of estimated values of both critical exponents between two Fibonacci subsequences of system sizes is negligible. Numerical analysis shows that the phase transition at $V_{c1}=2|\Delta-J|$ belongs to a different universality class from the quasiperiodic Ising universality at $V_{c2}$ with $\nu \simeq 1.000$, $z \simeq 1.388$~\cite{Lv2022Quantum}.

\section{Discussion and summary}\label{sec:SUMMARY}
In this paper, we pose an important and less understood question relating to the universality class of the transition from the extended phase to the critical phase in
the Aubry-Andr\'{e}-Harper (AAH) model with $p$-wave pairing, which are both gapped
under periodic boundary conditions.
Traditionally, for a conventional quantum many-body system described by a local Hamiltonian, it is 
accepted that two gapped
ground states that are connected without gap closing are
generally considered to belong to the same quantum phase. The undecidability of local order parameters exerts
an obstacle of comprehending the nature of the transition between the extended phase and the critical phase.
 Nevertheless, this phase transition is reflected in the spatial
variation of the wave function and can only be captured
by quantities that probe its extended or localized nature,
such as the inverse participation ratio (IPR), the bandwidth distribution and the level spacing distribution. To this end, we investigate the quantum criticality and universality in the AAH model from the perspective of information measures.
In particular, we can extract the universal information through the finite-size scaling of these measures, an impressive result given the limited number of system sizes in the quasiperiodic
systems, whose system sizes are rapidly growing three-subsequence Fibonacci numbers.

The extended phase and the critical phase are characterized by a logarithmic divergence of spectrum averaged entanglement entropy (SAEE) with the coefficients independent on the incommensurate modulation within each phase, in analogy to the bipartite entanglement entropy in the gapless phase. Notice that many-body entanglement entropy in two gapped phases admits the area-law scaling in this case.
In the extended state the prefactor is found to be equal to that of the conformally invariant homogeneous system, while in the critical state the prefactor is reduced to a smaller value, which can mark the fractal dimension. Meanwhile, the SAEE will saturate toward a constant for localized states.
It becomes evident that the
phase transitions in the AAH model are not thermodynamic phase transitions due to the absence of an explicit symmetry breaking. The transition between the extended phase and the critical phase can be ascribed to a reduction of the effective dimensionality across the critical point. We formulate Widom-like scaling ansatzs for the SAEE and its derivative, which are corroborated by an acceptable collapse for odd number of lattice sizes with a properly chosen correlation-length critical exponent $\nu$. The numerical results indicate that the multipartite entanglement can provide us a deep understanding of critical phenomena in quasiperiodic systems.
Furthermore, we show that the eigenstate generalized fidelity susceptibility (GFS)
proves to be an accelerated method for identifying the location of critical points.  More importantly, the versatility in GFSs of different orders
poses an efficient avenue for retrieving the dynamical critical exponent $z$, which is unable to be obtained from nonclosing gap and standard fidelity susceptibility across the extended-critical transition. The GFSs in the vicinity of the critical point $V_{c1}$ scale onto two separate scaling functions for each subsequence of odd system sizes $N$ with $\nu\simeq 1.000$ and $z\simeq 3.610$.
The extracted critical exponents are different from ones of the critical-localized transition point $V_{c2}$, i.e., $\nu \simeq 1.000$ and $z \simeq 1.388$.

Our work is interesting from various points of view. First, to the best of our knowledge, the critical exponents and the universal scaling analysis of the extended-critical transition in the $p$-wave-paired AAH model and other variants have not been retrieved yet. Second, we find that multipartite entanglement serves as a good indicator of phase transitions with interesting scaling behavior, which can recapitulate the fractal dimension. Third, we develop the strategy in terms of GFSs to analyze the critical phenomena in quasiperiodic systems and prove that the eigenstate GFS plays an irreplaceable role in describing the phase transition without gap closing. We emphasize the phase transitions under consideration here, that cannot be diagnosed by the quantities associated with many-body ground state, occur at the level of a single eigenstate in non-interacting systems. Last but not least, the rapid progress in quantum simulation sheds light on the experimental measurement of the correlation-length exponent $\nu$ and the dynamical exponent $z$ with a finite number of ultracold atoms. For instance, $z$ and $\nu$ may be extracted from specific-heat exponent $\alpha$ according to the hyperscaling relation $ (d+z) \nu $=$2-\alpha$ or through the Kibble-Zurek exponent $\mu=\nu/(1+\nu z)$~\cite{Sinha2019Kibble,chepiga2021kibble}.
Thus, the scaling hypothesis provides a unified and comprehensive description of universal order parameters, which is propitious for
extracting the associated universal information from limited system sizes of quasiperiodic quantum critical points. Our work paves a new routine of exploiting quantum criticality and universality in quasiperiodic and disordered systems. As a possible future direction, the present investigations could be extended to various generalized AAH models. We expect that the universality of the extended-critical transition may preserve and in particular the scaling analysis should hold in generalized AAH models, which even exhibit a mobility edge in the single-particle spectrum.

\section*{ACKNOWLEDGMENTS}
This work is supported by the National Natural Science Foundation of China (NSFC) under Grant No. 12174194, the startup fund of Nanjing University of Aeronautics and Astronautics under Grant No. 1008-YAH20006, Top-notch Academic Programs Project of Jiangsu Higher Education Institutions (TAPP), and stable supports for basic institute research under Grant No. 190101.

\appendix
\section{Ground-state fidelity susceptibility in free Fermi systems}
 \label{sec:GSFS}
\begin{figure}[tbh]
\centering
\includegraphics[width=\columnwidth]{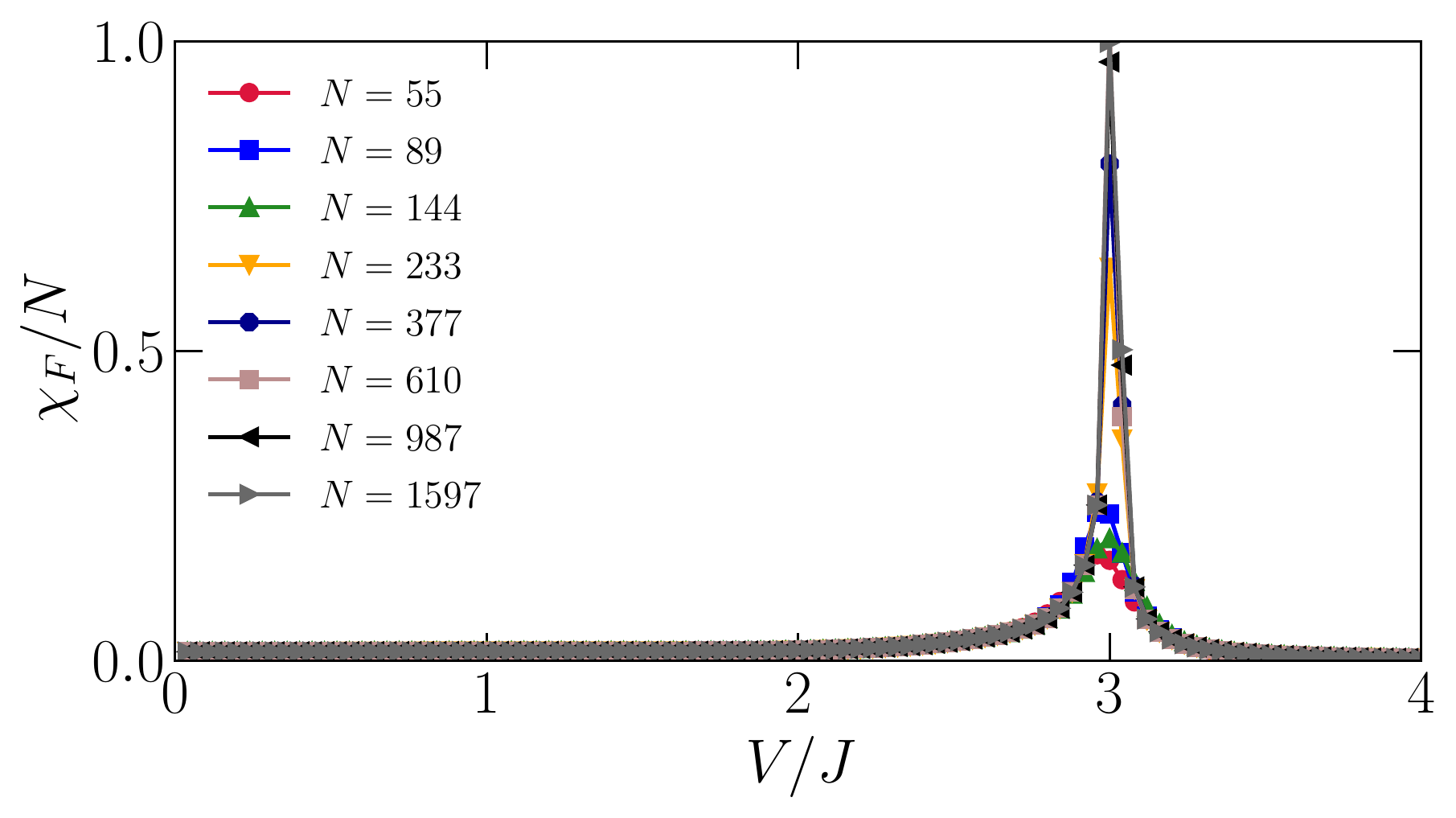}
 \caption{The ground-state fidelity susceptibility $\chi_{\rm F}/N$ as a function of the incommensurate potential strength $V$ for different sizes $N$. Here periodic boundary conditions are used with $\Delta=0.5$, $\phi=\pi$. }
  \label{fig:GSFS}
\end{figure}
Eq.(\ref{Ham}) describes a quadratic Hamiltonian for quasifree spinless fermions that in general is given by
\begin{equation}\label{eq:Hamfree}
H=\sum_{i,j=1}^{N} c^{\dagger}_i A_{ij}c_j +
\frac{1}{2} \sum_{i,j=1}^{N} \left( c^{\dagger}_i B_{ij}c^{\dagger}_j + c_j B_{ij} c_i\right),
\end{equation}
where $A$ ($B$) is a symmetric (antisymmetric) real $N \times N$ matrix,
i.e., $A^T=A,\, B^T=-B$.  The  Hamiltonian (\ref{eq:Hamfree}) can be rewritten in the more compact form as
$H=(\bm\Psi^\dagger C \bm\Psi+\text{Tr}\,A)/2$,
where  $\bm\Psi^\dagger=(c_1^\dagger,\dots,c_N^\dagger,c_1^{},\dots,c_N^{})$, $\bm\Psi=(c_1^{},\dots,c_N^{},c_1^\dagger,\dots,c_N^\dagger)^T$,
 and
$C=\sigma_z\otimes{}A+i\sigma_y\otimes{}B$.
In this case, the diagonalization of Eq.~(\ref{eq:Hamfree}) can be implemented efficiently, as it involves
operations in the Nambu space of dimension $N^2$, much less than the dimension $2^N$ of the full Hilbert space.

We consider a linear transformation $\bm\Psi'=V\bm\Psi$, where
$V=\openone_2\otimes{}u+\sigma_x\otimes{}v$ is orthogonal,
with $u$, $v$ being real matrices.
The transformation is canonical due to the  preservation of the anticommutation relations:
\begin{eqnarray}\label{eq:can_cond}
&&uu^T+vv^T=\openone\, \quad {\rm and} \quad  uv^T+(uv^T)^T=0\ ,
\end{eqnarray}
Under the real canonical transformations, the Hamiltonian becomes
$H=({\bm\Psi'}^\dagger{}C'\bm\Psi'+\mathrm{Tr}A)/2$, where
the eigenvalue matrix $C'=VCV^T=\sigma_z\otimes\Lambda$ with the $N\times{}N$  positive semi-definite diagonal matrix
$\Lambda=\mathrm{diag}(\Lambda_1,\dots,\Lambda_N)$.
Here the fact that the eigenvalues of $C$ appear in pairs of real numbers of
opposite sign is guaranteed by the imposed particle-hole symmetry in the Nambu representation.

In fact, all information of Eq.~(\ref{eq:Hamfree}), including ground-state
properties, can be decoded from a $N^2$-dimensional auxiliary real matrix $Z \equiv A - B$ \cite{Ground2007Zanardi,Scaling2009Jacobson,Fidelity2009Garnerone}.
 One can thus find $Z'=A'-B'$ satisfies
the simple relation
\begin{eqnarray}
\label{eq:Z'}
Z'=(u+v)Z(u-v)^T \ .
\end{eqnarray}
One then has $C'=\sigma_z\otimes{}A'+i\sigma_y\otimes{}B'$,
so that $A'=\Lambda$, $B'=0$ and hence $Z'=A'-B'=\Lambda$.
From Eq.~(\ref{eq:Z'}), by defining $\Phi\equiv{}u+v$ and $\Psi\equiv{}u-v$, one
finally gets the important equation
\begin{eqnarray}
\label{eq:PhiZPsi^T}
\Phi{}Z\Psi^T = \Lambda  \quad {\rm and} \quad \Psi{}Z^T\Phi^T=\Lambda \ .
\end{eqnarray}
Consequently, $\Phi$, $\Psi$, and $\Lambda$ are simply given
by the \emph{singular value decomposition} of $Z=\Phi^{T}\Lambda\Psi$.
Due to the canonical conditions,
the matrices
$\Phi$ and $\Psi$ must be orthogonal.
Then we can derive straightforwardly the following relations
$\Lambda\Psi=\Phi{}Z$,
$\Lambda\Phi=\Psi{}Z^T$.
As a consequence, $\Phi$ and $\Psi$  can be calculated by diagonalizing $ZZ^T$ or $Z^TZ$, given by
\begin{eqnarray}\label{eq:ZZT}
\Phi{}ZZ^T=\Lambda^2\Phi, \quad {\rm and} \quad
\Psi{}Z^TZ=\Lambda^2\Psi.
\end{eqnarray}
Note that $\Phi$ and $\Psi$  cannot be calculated by diagonalizing $ZZ^T$ and $Z^TZ$ independently because of Eq.~(\ref{eq:PhiZPsi^T}).
After solving Eq.~(\ref{eq:ZZT}) for orthogonal matrices $\Phi$, $\Psi$ and diagonal matrix $\Lambda$,  one introduces $u\equiv(\Phi+\Psi)/2$ and
$v\equiv(\Phi-\Psi)/2$, in terms of which the canonically transformed operators
diagonalizing the Hamiltonian are defined by
\begin{eqnarray}\eta_k\equiv\sum_{j=1}^N(u_{k,j}^{}c_j^{}+v_{k,j}^{}c_j^\dagger).
\end{eqnarray}
Finally, the Hamiltonian reads
\begin{eqnarray}
H = \sum_{k=1}^N\Lambda_k^{}\eta_k^\dagger\eta_k^{}+E_0 \ ,
\end{eqnarray}
where $E_0=\mathrm{Tr}(A-\Lambda)/2$ is the ground-state energy with $\Lambda_k$ being single particle energies.

The ground state can be obtained  by
imposing the  condition $\eta_k \ket{\Psi_0}=0,\; \forall \; k$
\cite{Peschel2003Calculation}. Recalling the \emph{singular value decomposition} of $Z$,
\begin{eqnarray}
Z=\Phi^{T}\Lambda\Psi=(\Phi^{T}\Lambda \Phi^{*})(\Phi^{T}\Psi)  = P T,
\end{eqnarray}
giving rise to the polar decomposition of $Z$
such that $Z=P T$, where $P \equiv  \sqrt{ZZ^\dagger} = \Phi^{T}\Lambda \Phi^{*} $ is a positive semi-definite
matrix and
$T=\Phi^{T}\Psi$ is unitary.
The ground state has an explicit BCS-like form in the case
of even parity \cite{Peschel2003Calculation}
\begin{eqnarray}
\label{Eq:Psi0}
\vert g_{Z} \rangle= \mathcal{N} \exp{\left( \frac{1}{2} \sum_{j,k=1}^{N} c_j^{\dagger} G_{jk} c_k^{\dagger} \right)}
\vert 0 \rangle,
\end{eqnarray}
where $ \mathcal{N} $ is a normalization factor, $\ket{0}$ is the fermionic vacuum ($c_j\ket{0}=0$) and
$G$ is a real $N\times N$ antisymmetric matrix, which satisfies
\begin{eqnarray}
\label{eq:gG+h}
uG+v=0 \ .
\end{eqnarray}
Note that Eq.(\ref{eq:gG+h}) is not always solvable, corresponding to cases where either the ansatz~(\ref{Eq:Psi0}) does not hold or the ground state parity is odd. We will put these exceptions on hold for the moment. We will assume $Z$ is invertible and $T$ is well defined in the following.

When $P$ (and hence $Z$) is invertible,
by using Eq.~(\ref{eq:PhiZPsi^T}) one can write
$u=\Phi(\openone+P_\Phi^{-1}Z)/2$,
$v=\Phi(\openone-P_\Phi^{-1}Z)/2$,
with $P_\Phi \equiv \Phi^{-1}P\Phi$,
so that if $u$ is invertible one has
\begin{eqnarray}
\label{eq:G}
G = \frac{T-\openone}{T+\openone} \ ,
\end{eqnarray}
where $G$ is the Cayley transform  of
$T\equiv P_\Phi^{-1}Z$, which is the orthogonal part of the
polar decomposition of $Z$. From Eq.~(\ref{eq:G}), the orthogonality $T^T=T^{-1}$ readily implies
the antisymmetry $G^T=-G$.
The inverse Cayley transform then yields  $T=(\openone+G)/(\openone-G)$,
implying $\det{T}=1$. We can then find that the spectrum of the real antisymmetric matrix $G$ is given by complex conjugate pairs of purely imaginary eigenvalues.

With the above wisdom, it is ready to calculate the ground-state fidelity as
\begin{eqnarray}\label{eq:fid0}
&&F(Z,\tilde{Z})  \equiv  \vert\langle g_Z \vert g_{\tilde{Z}}\rangle\vert.
\end{eqnarray}
We give an explicit evaluation of the fidelity (\ref{eq:fid0}) from the unitary matrix $T$, which can be further simplified into the following form by recognizing
that (\ref{Eq:Psi0}) is a fermionic coherent state \cite{Quantum2007Cozzini}:
\begin{eqnarray} \label{eq:fid}
F(Z,\tilde{Z})&=&\frac{ \det(\openone + G^\dagger \tilde G)^{1/2} }
{\det(\openone + G^\dagger G)^{1/4}\det(\openone + \tilde{G}^\dagger \tilde{G})^{1/4}}\nonumber \\
&=&\sqrt{\vert \det{\frac{\openone+T^{-1}\tilde{T}}{2}} \vert}=\sqrt{\vert \det{\frac{T+\tilde{T}}{2}} \vert}, \quad
\end{eqnarray}
where $T$ and $\tilde{T}$ are respectively
the unitary matrices of $Z \equiv Z(V)$ and $\tilde{Z} \equiv Z(V+\delta V)$ for infinitesimally close parameters $V$ and $V+\delta V$.

Using $\sqrt{\det(M)}=\det(\sqrt{M})$ and $\det(e^M)=e^{\textrm{Tr}(M)}$, Eq.(\ref{eq:fid}) can be further simplified as
\begin{eqnarray}\label{eq:derivation_1}
\lefteqn{F(Z,\tilde{Z})=
\sqrt{\vert \det{\frac{T+\tilde{T}}{2}} \vert}}\nonumber \\ &=&
\exp\left\{\textrm{Tr} \ln{\left(\frac{1+T^{\dagger}\tilde{T}}{2}\right)^{1/2}}\right\}\nonumber \\
&=&\exp\left\{\textrm{Tr} \ln{\left(\frac{1+T^{\dagger}\left(T+\delta T \right) }{2}\right)^{1/2}}\right\}\nonumber \\
&=&\exp\left\{\textrm{Tr} \ln{\left(1+\frac{T^{\dagger}\delta T}{2}\right)^{1/2}}\right\}\nonumber \\
&=&\exp\left\{\textrm{Tr} \frac{1}{2}\ln{\left(1+\frac{T^{\dagger}\delta T}{2}\right)}\right\}\nonumber \\
&\approx&\exp\left\{\textrm{Tr}   \left[ \frac{1}{4}(T^{\dagger}\delta T)-\frac{1}{16}\left( T^{\dagger}\delta T \right)^2 \right]\right\},
\end{eqnarray}
where $\delta T=\partial_{V}T d V $.
The fidelity susceptibility is thus obtained by \cite{You2007Fidelity}
\begin{equation}\label{eq:chi}
\chi_{\rm F} (V)=\lim_{\delta V \rightarrow 0} \frac{-2 \ln{F}}{\delta V^2}.
\end{equation}
To this end, Eq.(\ref{eq:chi}) can be rewritten in terms of the unitary matrix $T$ as
\begin{equation}\label{eq:chifrob}
\chi_{\rm F} =\frac{1}{8}\Vert\partial_{V}T \Vert_{F}^{2},
\end{equation}
where $\Vert M \Vert_{F}$$\equiv \sqrt{\textrm{Tr}(M M^{\dagger})}$ is the Frobenius norm. As is observed in Fig.\ref{fig:GSFS}, the fidelity susceptibility manifests divergent peaks at $V_{c2}$, while there is no anomaly across $V_{c1}$  without gap closing.

\section{Spectrum averaged fidelity susceptibility}
 \label{sec:SAFS}
\begin{figure}[tbh]
\centering
\includegraphics[width=\columnwidth]{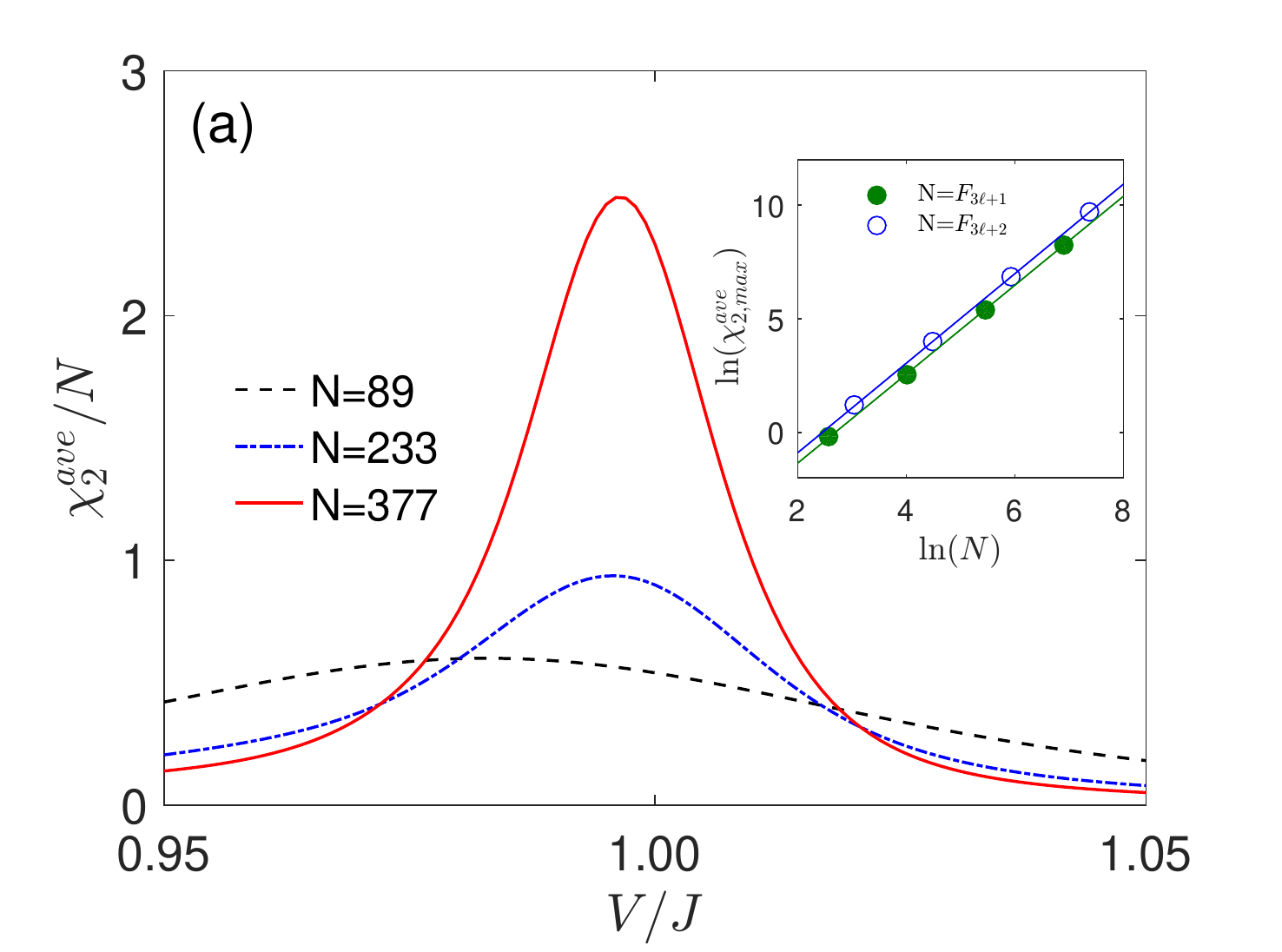}
\includegraphics[width=\columnwidth]{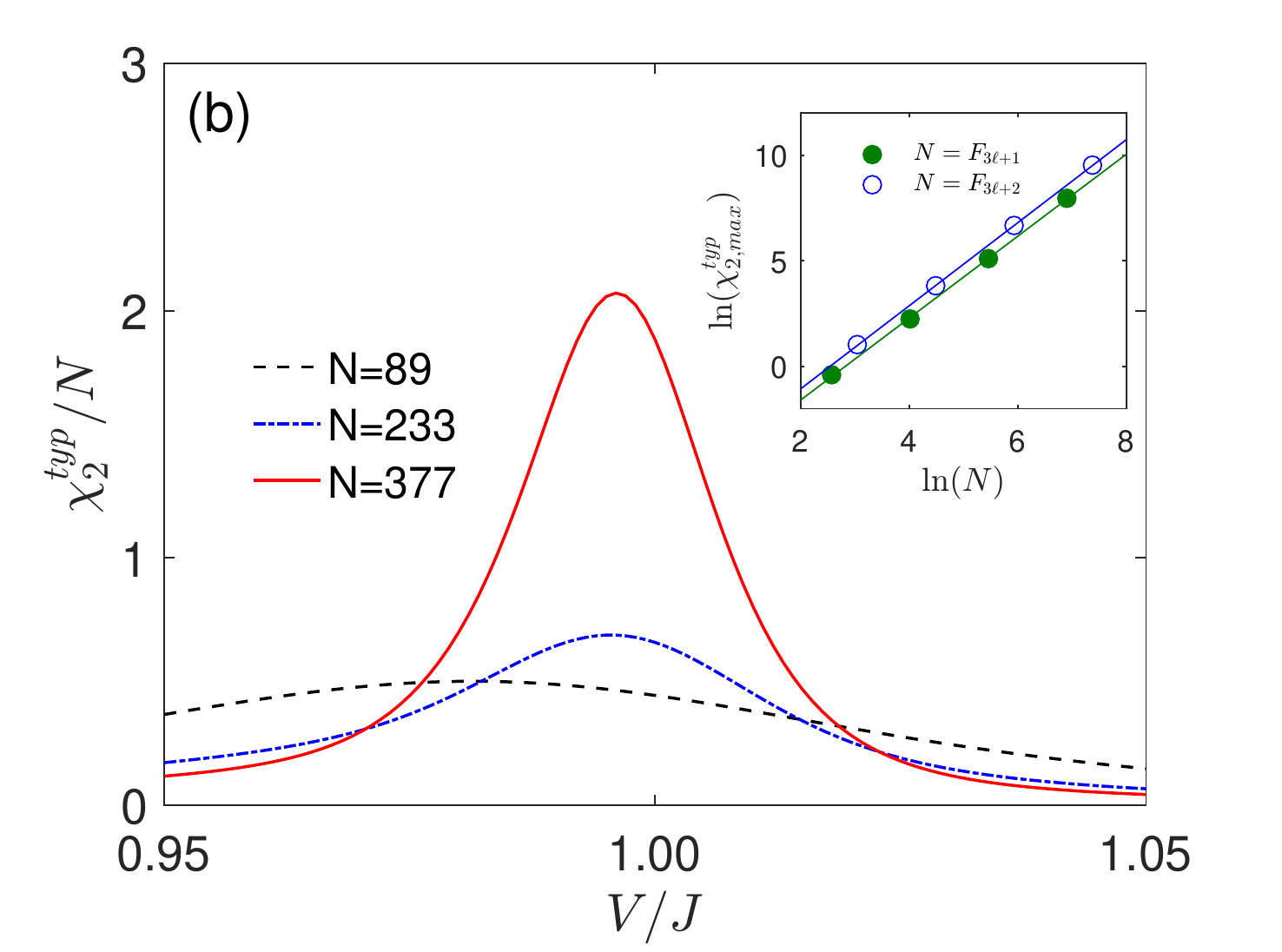}
 \caption{The spectrum averaged fidelity susceptibility per site (a) $\chi^{ave}_{2}/N$ and (b) $\chi^{typ}_{2}/N$ as a function of the incommensurate potential strength $V$ around $V_{c1}=1.0$. The inset shows the scaling behavior of the maxima versus the system sizes for $N=F_{3\ell+1}$: $13$, $55$, $233$, $987$,  and $N=F_{3\ell+2}$: $21$, $89$, $377$, $1597$.   Here periodic boundary conditions are used with $\Delta=0.5$, $\phi=0$. }
  \label{fig:SAFS}
\end{figure}
It has been widely recognized that the fidelity susceptibility is extensive far away from criticality and superextensive at criticality with a vanishing small gap.  In computing the spectrum averaged fidelity susceptibility, the arithmetic mean is highly sensitive to large values of certain eigenstates, while the geometric mean gives rise to a more representative measure of the typical values of the physical quantity under consideration. Especially in the vicinity of the critical points, large fidelity susceptibilities
skew the arithmetic average toward a greater value than the typical one. The arithmetic mean generally provides an upper bound for the geometric mean when taking the mean of a set of positive values and they become equal only when averaging over a constant set of values. We will define the average fidelity susceptibility as
\begin{eqnarray}
\label{eq:ave}
\chi_{2}^{\rm ave}\equiv \frac{1}{2N}\sum_{k=1}^{2N} \chi^{(k)}_{2},
\end{eqnarray}
and the typical fidelity susceptibility as
\begin{eqnarray}
\label{eq:typ}
\chi_{2}^{\rm typ}\equiv  \exp\left( \frac{1}{2N}\sum_{k=1}^{2N}\ln \chi^{(k)}_{2}\right).
\end{eqnarray}
We show both types of spectrum averaged fidelity susceptibilities
for the AAH model with $p$-wave pairing around $V_{c1}=1.0$. We can observe similar behaviors in Fig.\ref{fig:SAFS}.
Linear fits of local peaks reveal
$\ln(\chi_{2,{\rm max}}^{typ} )$ = $(1.941 \pm 0.121)$  $\ln N$-$(5.472\pm 0.608)$ and $\ln(\chi_{2,{\rm max}}^{ave})$ = $(1.957  \pm 0.038)$  $\ln N$-$(5.260\pm 0.388)$ for $N=F_{3\ell+1}$, while for $N=F_{3\ell+2}$, we have $\ln(\chi_{2,{\rm max}}^{typ})$ = $(1.9678 \pm 0.054)$  $\ln N$-$(4.994\pm 0.295)$
and $\ln(\chi_{2,{\rm max}}^{ave})$ = $(1.969 \pm 0.048)$  $\ln N$-$(4.827\pm 0.265)$, suggesting that the extracted coefficients of the slopes for the correlation-length exponent agree well with each other.
It is obvious that there is no qualitative difference
between the spectrum averaged fidelity susceptibilities [cf. Fig.\ref{fig:SAFS}] and the fidelity susceptibility of $\vert\psi_1\rangle $ [cf. Fig.\ref{fig:AAHmodelfs2}(a)]. Therefore, it is sufficient to retrieve the universal information through the finite-size scaling of the lowest eigenstate for systems without mobility edges.

\bibliography{references_0910}
\end{document}